\documentclass[letterpaper]{article} 

\usepackage{aaai2026}
\usepackage{times}  
\usepackage{helvet}  
\usepackage{courier}  
\usepackage[hyphens]{url}  
\usepackage{graphicx} 
\usepackage{natbib}  
\usepackage{caption} 
\usepackage{xspace}
\usepackage{booktabs} 
\usepackage{xcolor} 
\usepackage{xspace}
\usepackage{booktabs} 
\usepackage{graphicx} 
\usepackage{caption}
\usepackage{subcaption}
\usepackage[labelfont=bf]{caption}
\usepackage{tikz}
\usepackage{makecell}
\usepackage{multicol}

\usepackage{tabularx}
\usepackage{booktabs}
\usepackage{array}

\usepackage{makecell}  

\usepackage[table]{xcolor}

\newcolumntype{Y}{>{\raggedright\arraybackslash}X}
\newcolumntype{Z}{>{\centering\arraybackslash}m{0.8cm}} 
\newcolumntype{C}[1]{>{\centering\arraybackslash}m{#1}}  

\definecolor{highcolor}{HTML}{EDF7EB}   
\definecolor{mediumcolor}{HTML}{FFF9E5} 
\definecolor{lowcolor}{HTML}{FCECEC}    

\definecolor{lightpurple}{HTML}{D6CCF2} 
\definecolor{lightgreen}{HTML}{D9EAD3}
\definecolor{lightblue}{HTML}{CFE2F3}
\definecolor{lightyellow}{HTML}{FFF2CC}

\definecolor{lightpink}{HTML}{EAD1DC}
\definecolor{lilac}{HTML}{D8B7DD}
\definecolor{lightgrey}{HTML}{F0F0F0}

\newcommand{\stepcircled}[1]{%
  \tikz[baseline=(char.base)]{
    \node[shape=circle,draw=black,fill=black,inner sep=1pt] (char) {\textcolor{white}{\textbf{#1}}};
  }%
}

\usepackage{newfloat}
\usepackage{listings}

\usepackage{algorithm}
\usepackage{algorithmic}
\usepackage{pdfpages}
\usepackage{subcaption}
\usepackage{caption}
\usepackage{subcaption}

\usepackage{xspace}
\usepackage{longtable}
\usepackage{booktabs}
\usepackage{multirow}

\newcommand*{\eg}{e.g.,\xspace}
\newcommand*{\ie}{i.e.,\xspace}

\newcommand*\framework{\emph{Co-PALE}\xspace}
\newcommand*\frameworklong{\emph{Contextualized Perceptions for the Adoption of LLMs in Education}\xspace}

\newcommand{\submit}[1]{}

\title{``Would You Want an AI Tutor?''  Understanding Stakeholder Perceptions of LLM-based Systems in the Classroom}

\author{
    Caterina Fuligni\textsuperscript{\rm 1},
    Daniel Dominguez Figaredo\textsuperscript{\rm 2},
    Armanda Lewis\textsuperscript{\rm 1},
    Julia Stoyanovich\textsuperscript{\rm 1}
}
\affiliations{
    \textsuperscript{\rm 1}New York University, New York City, New York, USA\\
    \textsuperscript{\rm 2}Universidad Nacional de Educación a Distancia, Madrid, Spain\\
    cf3181@nyu.edu, ddominguez@edu.uned.es, al861@nyu.edu, stoyanovich@nyu.edu
}

\nocopyright

\begin{document}

\maketitle

\begin{abstract}
Large Language Models (LLMs) have gained traction in educational settings, often framed as virtual tutors or teaching assistants. Following early skepticism and bans, many schools and universities have begun integrating these systems into curricula. Yet decisions about whether and how to deploy LLM-based tools are frequently made without systematic engagement with the full range of stakeholders they affect.  In this paper, we argue that understanding stakeholder perceptions of LLM-based systems in the classroom is not a matter of measuring approval or acceptance, but of identifying whose concerns are surfaced, in which contexts, and with what implications for responsible design and governance. We introduce Contextualized Perceptions for the Adoption of LLMs in Education (Co-PALE), a stakeholder-first framework that connects educational context, responsible AI principles, and categories of perception to support more deliberate decision-making about the adoption of LLM-based tools.

We ground Co-PALE through a targeted analysis of prior work to diagnose recurring gaps in how stakeholder perceptions are studied, and through contextually distinct educational scenarios that illustrate how the same technology raises different concerns for different stakeholders. We further examine how university faculty and K--12 parents make sense of the framework through focus groups, using their reflections to surface tensions and uncertainties. Co-PALE supports more systematic reasoning about whether, where, and for whom LLM-based tools should be deployed in education.
\end{abstract}

\section{Introduction}
\label{sec:intro}

Large language models (LLMs) have spread rapidly across many sectors, including education. Trained on large-scale text corpora, LLMs can generate fluent natural language and exhibit contextual sensitivity, making them appealing across a range
of educational uses~\cite{watson2024chatgpt}.

LLM-powered tools present both opportunities and challenges for educational institutions, prompting schools and universities to take positions on their use~\cite{woodruff2023perceptions}. In the United States, school districts initially responded to the release of ChatGPT with bans that were later reversed~\cite{Rosenblatt.2023}. Since then, LLM-based tools have increasingly been integrated into curricula. For example, in 2023 more than 40 districts and 28{,}000 students and teachers piloted Khanmigo, Khan Academy's LLM-powered tutoring and teaching assistant system~\cite{Yamkovenko.2024}. Higher education institutions have likewise adopted AI tools to support instructional design, personalized learning, and administrative decision-making~\cite{khairullah2025implementing,ozfidan2024use, rahiman2024revolutionizing}. While some have embraced these technologies with enthusiasm, at times likening ChatGPT to the introduction of calculators or personal computers~\cite{shoufan2023exploring}, their adoption has also elicited concern and ambivalence, particularly among students and teachers~\cite{Engle.2023, Ravaglia.2024}.

\subsection{Motivating example}
\label{sec:intro:example}

In early 2023, OpenAI released GPT-4. As a launch partner, Khan Academy gained early access to the model and soon introduced Khanmigo, a GPT-4-powered tutoring and teaching assistant, through a limited pilot. Newark Public Schools was among the first U.S. districts to test the tool in classrooms~\cite{Yamkovenko.2024}.

Similar initiatives soon followed in New York City. The NYC Department of Education (DOE) partnered with Microsoft to pilot a custom AI-powered teaching assistant designed to answer students' questions and provide real-time feedback, initially in high school computer science courses. The tool was later extended with Microsoft's Math Solver and deployed in 15 public schools~\cite{Donaldson.2023}. At the same time, teachers in Northern Brooklyn were trained to use YourWai, an AI teaching assistant developed by Learning Innovation Catalyst, which was later criticized for displaying fabricated reviews on its website~\cite{bardolf.2023}. In 2024, NYC schools withdrew a \$1.9 million proposal for an AI-based reading tutor, Amira, following intervention by the city Comptroller over unresolved concerns regarding student privacy and governance. The tool had reportedly been used by more than 46{,}000 students across 162 NYC public schools~\cite{zimmerman.2024}.

These deployments unfolded amid limited public discussion of whether, how, and for whom such tools should be used. In an article titled ``Would You Want an A.I. Tutor?'', \emph{The New York Times} invited students to comment on their experiences with Khanmigo~\cite{Engle.2023}.\footnote{The title of our paper directly references~\citet{Engle.2023} that inspired our work.} Student responses revealed divergent perceptions. Some expressed concern about diminished human interaction, over-reliance on automated assistance, and uncertainty about accuracy, while others emphasized the system's accessibility, breadth of knowledge, and potential for personalized support.

Recent institutional guidance illustrates the limits of current approaches. The NYC Department of Education's March 2026 AI guidance adopts a categorical ``traffic light'' framework---green, yellow, and red use cases---that applies uniform rules across educational levels, disciplines, and stakeholder groups, without differentiating by who is affected or in what context~\cite{nycdoe2026}. Stakeholder perceptions are not treated as structured inputs to policy design; critics have noted that the guidance makes ``no attempt to address many of the most serious concerns that parents and educators have''~\cite{parentcoalition2026}. Even as institutional guidance evolves, it remains inattentive to the diversity of stakeholders, contexts, and perceptions at stake.

Related tensions have also emerged in higher education. As universities increasingly integrate generative AI into teaching practices, debates have intensified over transparency, academic integrity, and the boundaries of acceptable use. A May 2025 report describes a case in which a student at Northeastern University filed a formal complaint after discovering that course materials appeared to have been generated using ChatGPT, despite a syllabus prohibiting unauthorized AI use~\cite{Hill.2025}. The student argued that she was paying for instruction from human educators rather than algorithmically generated content; the complaint was ultimately denied.

Together, these cases illustrate a recurring pattern: LLM-based tools are introduced across educational contexts through pilot programs, partnerships, and procurement decisions, while the perceptions and concerns of affected stakeholders---students, families, educators, and administrators---are surfaced only informally, if at all, and often after deployment.

The position we take in this paper is that \emph{stakeholders}---individuals or groups directly or indirectly affected by the adoption of technology---should play an integral role in the design, development, implementation, and oversight of AI in socially salient contexts. Concerns about limited stakeholder involvement in educational AI decisions have been raised repeatedly~\cite{zeide2019robot}: school procurement and implementation are rarely part of public discussion, leaving students, families, and educators with little visibility into decisions that shape educational outcomes. A growing body of work calls for stakeholder-first approaches to AI development and governance, alongside stronger roles for civil society in establishing accountability mechanisms~\cite{fukuda2021emerging, bell2023think}.

\begin{figure*}
    \centering
    \includegraphics[width=0.85\linewidth]{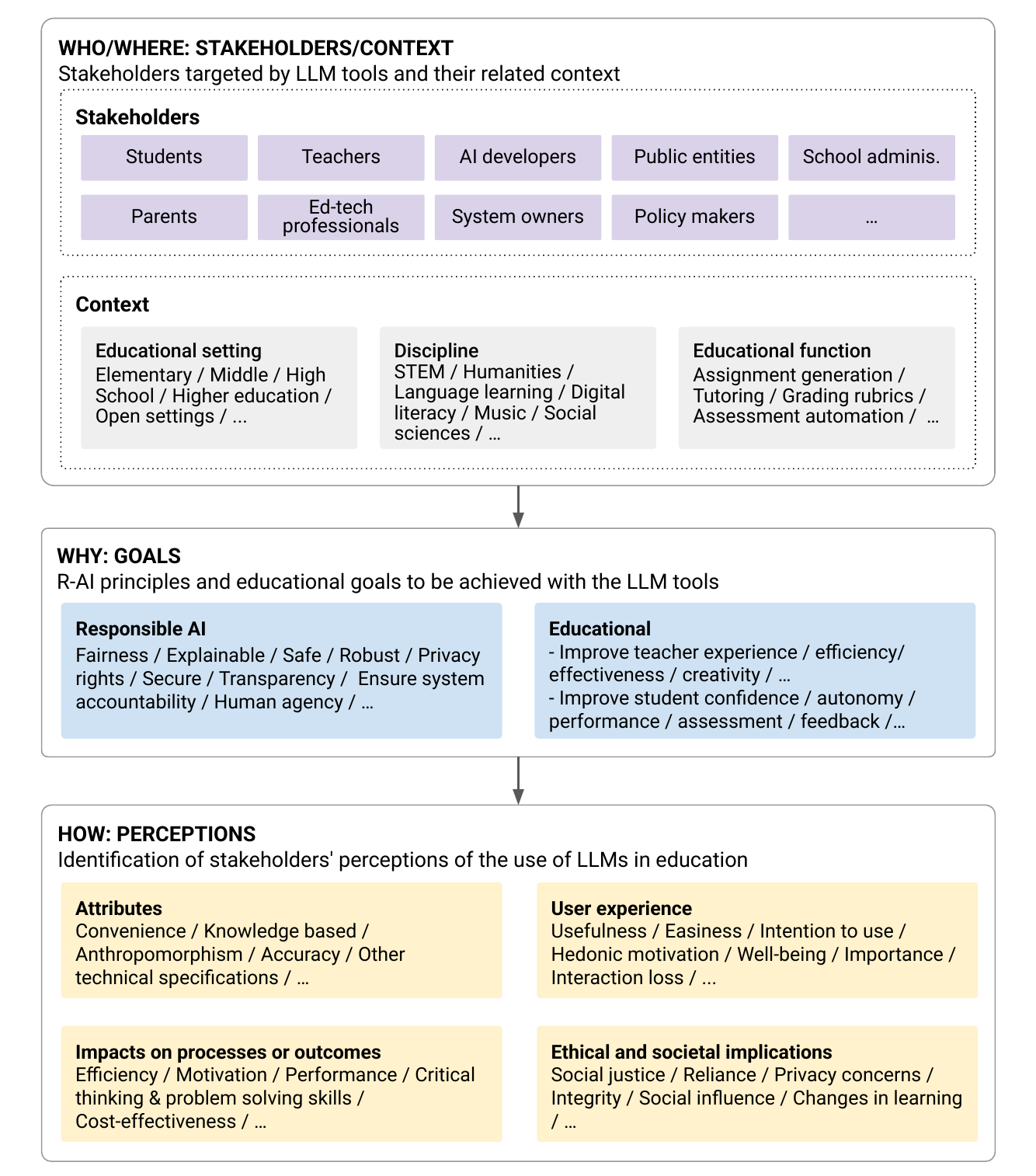}
    \vspace{-0.2cm}
    \caption{\frameworklong (\framework) framework for assessing stakeholder perceptions of LLM-based systems. \framework comprises four components: Stakeholders, Context, Goals, and Perceptions. Components should be considered sequentially, starting at the top and proceeding layer by layer. The Perceptions component can be used as a checklist to assess relevant perceptions, refine them, and add any that may be missing.}
    \vspace{-0.5cm}
    \label{fig:framework}
\end{figure*}

\subsection{Contributions}
\label{sec:intro:contrib}

This paper makes the following contributions toward a stakeholder-first, context-sensitive understanding of perceptions of LLM-based systems in educational settings.

\smallskip\textbf{\stepcircled{1}: We introduce the \frameworklong (\framework) framework} for reasoning systematically about stakeholder perceptions in relation to educational context and responsible AI goals. \framework connects four elements: stakeholders, context, goals, and a structured taxonomy of perception categories. The framework contextualizes perceptions by making explicit how concerns and priorities depend on who is affected and in what educational setting. \framework is summarized in Figure~\ref{fig:framework} and described in
Section~\ref{sec:framework}.

\smallskip\textbf{\stepcircled{2}: We demonstrate how \framework can be applied across distinct educational settings through contextually grounded scenarios.} These applications show how the same class of LLM-based tools can raise different and sometimes competing concerns for stakeholders depending on educational level, discipline, and instructional function. By foregrounding these contrasts, the scenarios illustrate how \framework can support more deliberate design, deployment, and governance decisions prior to and during adoption. Applications are presented in Section~\ref{sec:putting}.

\smallskip\textbf{\stepcircled{3}: We examine how two stakeholder groups, university faculty and K--12 parents, engage with \framework in practice through focus groups conducted across three international universities and the United States.} We conducted five in-person focus groups with faculty ($N=27$) and three online focus groups with parents ($N=24$). This empirical material is not intended as a comprehensive validation of the framework. Instead, it serves as a stress test that reveals how diverse stakeholders reason about perceptions under conditions of uncertainty, institutional constraint, and evolving norms around AI use in education. The focus group study and findings are reported in Section~\ref{sec:validation}.

\section{Related Work}
\label{sec:related}

\subsection{Stakeholder perceptions of LLM-based tools in education}
\label{sec:related:perceptions}

A growing body of work examines perceptions of LLM-based chatbots and tutors in educational settings using surveys, interviews, social media analysis, and related qualitative methods. Across studies, researchers document a mix of perceived opportunities and concerns, including effects on learning, productivity, academic integrity, and social interaction. At the same time, variation in stakeholder focus, educational context, and conceptualization of perceptions complicates synthesis and limits the applicability of these findings to concrete deployment decisions.

Early large-scale analyses illustrate this ambivalence. \citet{leiter2024chatgpt} analyzed over 300{,}000 tweets and more than 150 scientific papers published shortly after ChatGPT's release, finding a roughly balanced framing of the technology as both an educational opportunity and a potential threat, particularly with respect to academic integrity. Similarly, \citet{mogavi2024chatgpt} examined early user discussions across multiple social media platforms, documenting widespread use across higher education, K--12, and practical skills training, alongside tensions between perceived benefits such as productivity and motivation and concerns about overreliance and shallow learning.

\begin{figure*}[]
  \includegraphics[width=0.9\linewidth]{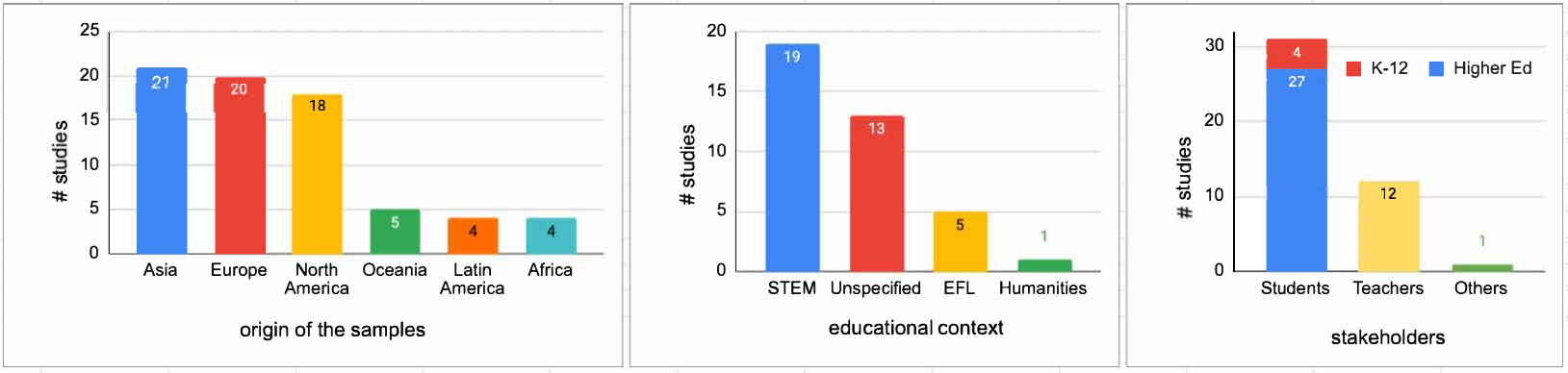}
  \caption{Breakdown of the analyzed articles ($N=38$) by geography, educational contexts, and stakeholder groups.}
  \label{fig:stats}
\end{figure*}

More focused empirical studies report similar patterns within narrower settings. \citet{dube2024students} examined university students' perceptions of ChatGPT in academic activities, identifying both perceived learning benefits and concerns related to creativity, trust, privacy, and academic misconduct. \citet{lo2023impact} analyzed ChatGPT's use across educational functions, finding variable performance across subjects and tasks and recurring concerns about accuracy, reliability, and plagiarism, even when the system was used for lesson preparation, assessment support, or tutoring. Work centered on K--12 education reflects similar ambivalence from the perspective of educators. \citet{kashif2025teachers} conducted a systematic review of teachers' views on AI integration in K--12 settings, finding optimism about potential instructional benefits and caution regarding associated risks.

Taken together, this literature offers valuable insight into early perceptions of LLM-based tools in education, while also suggesting areas of tension and inconsistency. Across studies, attention is concentrated on a narrow set of stakeholders, most often university students; educational contexts are described at a high level; and perceptions are measured using diverse frameworks and terminologies. 

To examine whether and how these patterns hold at scale, we conducted a \textit{targeted diagnostic} review of this literature---not a systematic review---using Google Scholar. The review was scoped to surface recurring gaps in stakeholder representation, contextual specificity, and perception assessment that would directly inform our framework's design, guided by two research questions: \emph{Who are the key stakeholders and what are their perceptions as identified in the literature? How are these perceptions
contextualized?} We included empirical studies that (i)~report data on stakeholder perceptions of LLM-based tools in education, (ii)~are situated in a formal educational context, (iii)~are available as full-text, peer-reviewed publications, and (iv)~are written in English. We excluded non-empirical work,
position papers, and studies focused primarily on tool performance rather than stakeholder response. Applying these criteria, we selected 38 articles for analysis. We developed a codebook through iterative readings, combining deductive and inductive approaches to code perceptions by sentiment (positive, negative, neutral), frequency, and stakeholder group; full search and coding details are provided in the Appendix.

Figure~\ref{fig:stats} gives the breakdown of the analyzed articles by geographic distribution, educational contexts, and stakeholder groups. Most studies included participants from Asia, Europe, and North America, with some drawing from multiple continents; when participant location was unspecified, we used the authors' institutional affiliation as a proxy. The most common educational context was STEM (50\%), followed by English as a foreign language (EFL) and Humanities, while over a third of the studies did not specify a context. As shown in Figure~\ref{fig:stats} (right), students were the most studied group (82\% of studies), followed by teachers (32\%), while groups such as parents or administrators were absent. We identified positive, negative, and neutral perceptions across four categories: \textit{Attributes}, \textit{Impact on Processes or Outcomes}, \textit{User Experience}, and \textit{Ethical and Societal Implications} (see Figure~\ref{fig:perceptions} in the Appendix). Student perceptions were predominantly positive (56\%), teacher perceptions were more often negative (52\%), and both groups expressed ethical concerns.

Our diagnostic analysis identified \textbf{three systematic gaps}, each of which directly shaped a core dimension of our framework (Section~\ref{sec:framework}). \textbf{First}, stakeholders other than students and teachers (\eg parents, school staff) are not represented in the literature; this motivated our expanded
stakeholder taxonomy, which explicitly includes indirect actors such as parents, school administrators, and ed-tech professionals. \textbf{Second}, educational contexts are narrow, with a strong focus on university-level STEM and limited attention to K--12, where students have less agency over technology use and thus require distinct consideration; this motivated our framework's context dimension, which distinguishes educational level, discipline, and instructional function. \textbf{Third}, assessment of stakeholder perceptions lacks consistency, with studies employing varied theoretical frameworks, terminologies, and measurement scales; this motivated our structured perception taxonomy---Attributes, User
Experience, Impact on Processes/Outcomes, and Ethical and Societal Implications---as a unifying classification intended to support comparability across settings.

\subsection{Stakeholder-centered approaches to AI in education}
\label{sec:related:frameworks}

Several studies have examined the role of stakeholders in the development and implementation of AI-powered tools in educational settings. \citet{chaudhry2022transparency} mapped transparency requirements for AI-based ed-tech tools by engaging educators, parents, ed-tech experts, and AI practitioners. They showed that transparency needs vary across stakeholder groups and evolve over time, and co-designed a Transparency Index, which they applied and iteratively refined in a real-world setting involving AI-powered educational tools developed for a training organization.

\citet{luckin2022empowering} proposed a seven-step AI readiness framework for the education and training sector. They argue that individuals and organizations achieve AI readiness by identifying organizational challenges best addressed by data and AI, the context for effective AI application, the goals of AI-based
tools, and the stakeholders involved. The framework emphasizes participatory processes that support educators and institutions in engaging with AI technologies.

\citet{garcia2024review} conducted a literature review of LLM-based solutions for generating and evaluating educational materials that involved students or teachers in their design or evaluation. They focused on promoting collaboration between researchers, developers, and end users, and found that LLM-based systems were most commonly used as virtual assistants for tasks such as question generation, answer grading, and code correction or explanation.

These studies illustrate a range of approaches to incorporating stakeholder considerations into the design and deployment of AI-based systems in education, differing in scope, stakeholder focus, and degree of contextualization. While they provide important foundations, none fully accounts for the diversity of
educational contexts, instructional functions, and stakeholder perceptions associated with LLM-based tools---the gaps identified in Section~\ref{sec:related:perceptions} that our framework is designed to address.

\section{Framework Design}
\label{sec:framework}

\subsection{Design rationale}
\label{sec:rationale}

A growing body of work at the intersection of responsible AI and education advocates for human-centered and stakeholder-aware approaches to the design and governance of AI systems. Table~\ref{tab:comparative} summarizes several such frameworks that inform the design of \framework, spanning user-centered explainable AI~\cite{meske2022explainable}, ethical governance practices~\cite{wang2020toward}, and human-centered documentation methodologies~\cite{richards2021human}.

Several frameworks explicitly adopt a stakeholder-first orientation. \citet{bell2023think} propose a stakeholder-first approach to address limitations of AI governance documents that specify high-level transparency requirements without operational guidance. Their framework begins by identifying relevant stakeholders and supports the design of transparent and compliant systems through iterative refinement. In the context of AI literacy, \citet{dominguez2023responsible} similarly argue that centering initiatives on intended stakeholders and tailoring content to their needs is essential for enabling critical engagement with algorithmic systems. \citet{baber2025human} emphasize collaboration among professionals, affected individuals, and policymakers to support ethical AI use, an approach that can be extended to education.

\begin{table*}[ht]
\centering
\small 
\caption{Comparative analysis of stakeholder-centered frameworks for responsible AI (RAI)}
\vspace{-0.2cm}
\resizebox{\textwidth}{!}{%
\begin{tabular}{p{3.8cm} p{3.6cm} p{3.6cm} p{4.2cm}}
\toprule
\textbf{Framework} & \textbf{Who} & \textbf{What} & \textbf{How} \\
\midrule
Explainable AI framework for transparency and trust (Meske et al., 2020) & Developers, Decision-makers, Consumers, Regulators & Transparency, Trust, Accountability through explainability & User-centric explainable models, enabling scrutiny, stakeholder collaboration  \\
\midrule
A framework for RAI practices in companies (Wang et al., 2020) & Corporate stakeholders, Employees, End-users & Data governance, ethical design solutions, human-centered oversight, training and education & Stakeholder involvement, policy implementation, risk assessment and mitigation \\
\midrule
Human-centered methodology for creating AI FactSheets (Richards et al., 2021) & AI developers, End-users, Auditors & Transparency, accountability, user understanding of AI systems & Iterative human-centered design, stakeholder feedback, structured documentation \\
\midrule
A stakeholder-first approach framework for creating transparent ADS (Bell et al., 2023) & Technologists, Policymakers, Auditors, Regulators, Humans-in-the-loop, Affected individuals & RAI principles (e.g., trust, fairness, privacy), transparency goals, developers’ purposes & Methods for designing compliant systems, stakeholder consultation, iterative refinement \\
\midrule
A stakeholder-first approach framework for designing AI literacy practices (Domínguez \& Stoyanovich, 2023) & System owners, public entities, civil society. 
& Include the term/principles of ``responsibility'' throughout the content design cycle & Tailored learning resources: adapted to different levels of difficulty and learning situations; customized strategies by audience; contextual adaptation \\
\midrule
RAI framework for innovation in medicine (Baber et al., 2025) & Healthcare professionals, Patients, Policy makers, Developers & Ethical AI use, fairness, accountability, risk minimization & Collaboration, ethical guidelines, continuous monitoring \\
\bottomrule
\end{tabular}
}
\vspace{-0.2cm}
\label{tab:comparative}
\end{table*}

While these approaches provide important foundations, they do not fully account for the diversity of educational contexts, instructional functions, or stakeholder perceptions associated with LLM-based tools. \framework builds on this work by integrating three considerations central to educational AI adoption: identifying direct and indirect stakeholders, specifying educational context, and supporting structured reasoning about stakeholder perceptions in relation to responsible AI principles and educational goals. \framework identifies relevant stakeholders and context, including educational setting, discipline, and intended function, and supports the systematic assessment of salient perceptions through a structured workflow for reasoning about design, deployment, and governance decisions (Figure~\ref{fig:framework}).

Unlike prevailing educational AI governance approaches, which often function as retrospective compliance checklists or pass-fail capability assessments rather than forward-looking deliberative tools~\cite{yin2025responsible,kallina2025stakeholder}, \framework is designed as a procedural workflow. Rather than auditing systems against abstract benchmarks after deployment, it guides structured reasoning about stakeholder perceptions before and during adoption decisions.

\subsection{Structure and layers}
\label{sec:framework:structureandlayers}
\framework provides a structured system for integrating stakeholder perceptions into decisions about the design, deployment, and governance of LLM-based technologies in educational settings. The framework's elements are intentionally open, allowing users to adapt and extend them to match specific application contexts.
\framework comprises four dimensions:
\emph{stakeholders}, \emph{context}, \emph{goals}, and \emph{taxonomy of perceptions}. Starting at the top, elements of each dimension must be considered before moving on to the next one. First, the framework identifies the stakeholders who may interact with the AI systems. Then, it outlines a workflow that incorporates the key elements needed to identify perceptions within specific learning contexts. The final dimension focuses on perceptions, and offers scales and rubrics for assessing them.

\smallskip\textbf{Stakeholders.} We highlighted how the introduction of LLM-based tools, such as Khanmigo, in the classroom has followed a top-down decision-making process, where partnerships between private companies and governmental institutions or school districts often make decisions on behalf of direct stakeholders, such as teachers and students, or indirect stakeholders, such as parents and school staff. This echoes a concern already highlighted in the literature regarding how private companies, more often than community members and public servants, set pedagogy and policy in practice \cite{zeide2019robot}. To counteract this, our framework prioritizes educational stakeholders: we identified students and teachers as the most recurring stakeholders in the literature; however, as shown in Section~\ref{sec:related:perceptions}, existing empirical work largely omits other key stakeholders, such as parents, school staff, ed-tech professionals, and government agencies.

\smallskip\textbf{Context.} As discussed in Section~\ref{sec:related:perceptions}, most studies in the literature focus on higher education. \framework therefore distinguishes educational levels, including contexts involving younger students who may have limited autonomy over technology use. The literature also shows a strong emphasis on STEM, with many studies leaving application areas unspecified, motivating explicit differentiation among disciplines. This section considers three levels of specificity: the \emph{educational setting} (\ie educational level), the \emph{discipline} (\eg STEM, language learning, humanities), and the \emph{educational function} for which the tool is used (\eg assignment generation, tutoring, assessment).

\smallskip\textbf{Goals.} \framework draws on two types of goals. \textit{Responsible AI goals} include fairness, explainability, transparency, safety, robustness, privacy, accountability, and human agency~\cite{bell2023think, chaudhry2022transparency, Murad2022}. \textit{Pedagogical goals} focus on embedding these principles in educational practice and linking them to specific learning outcomes~\cite{dominguez2020data, fu2024navigating}, for example through adaptive feedback and grounded materials~\cite{jurenka2024towards}, and participatory design that supports learner ownership~\cite{louie2022designing}.

\smallskip\textbf{Perceptions.} To support more consistent reasoning about stakeholder perceptions, \framework proposes a high-level, open-ended categorization of perception types, intended as a checklist for practitioners and researchers reasoning about AI-based tool adoption in education. Perception types are grouped into four categories: Attributes, User Experience, Impact on Processes/Outcomes, and Ethical and Societal Implications. The taxonomy aims to unify terminology across the literature and is designed to be refined through practical application.

\section{Putting \framework into practical use}
\label{sec:putting}

We illustrate how to apply \framework by operationalizing the framework through an iterative, three-step workflow inspired by \citet{homewood2025third}. \textbf{First}, practitioners must understand the unique contours of their educational setting to
logically evaluate vulnerabilities and formulate governance guidelines. \textbf{Second}, we provide a rubric listing all cases contained in \framework resulting from the intersection of stakeholders and perceptions. Each case is labeled as high, medium, or
low importance. \textbf{Third}, users refer to the labeled map when making decisions about the AI tool deployed in their educational setting. A wide range of educational stakeholders, whether directly or indirectly involved in using LLM-based tools, can apply the framework and use the rubric to label their perceptions on the resulting map.

\subsection{Mapping the learning situation}
\label{sec:putting:mapping}

The initial mapping phase establishes the contextual baseline necessary for any granular evaluation. The workflow begins with users identifying the contextual parameters specific to their educational situation. Table~\ref{tab:template} provides a standardized template for this phase, guiding users to map the boundaries of a proposed AI deployment along three descriptive axes representing the ``who,'' ``where,'' and ``why'' of the learning scenario. First, users identify the stakeholders, documenting direct participants, such as students and teachers, as well as critical indirect actors, including parents, school administrators, and ed-tech professionals, whose roles may change with tool adoption. Next, users establish the institutional and situational context by defining parameters such as the educational level, academic discipline, and the tool's precise functional role within the classroom. Finally, users articulate their primary objectives, balancing pedagogical goals with responsible AI principles.

\begin{table*}[h!]
\small 
\centering
\caption{\framework mapping template}
\vspace{-0.3cm}
\label{tab:template}

\begin{tabularx}{\textwidth}{@{}>{\columncolor{lightpurple}\bfseries}Y
                            >{\columncolor{lightgrey}\bfseries}Y
                            >{\columncolor{lightblue}\bfseries}Y
                            >{\columncolor{lightblue}\bfseries}Y@{}}
\toprule
\rule{0pt}{2.5ex}Stakeholder & \rule{0pt}{2.5ex}Context & \rule{0pt}{2.5ex}R-AI Goals & \rule{0pt}{2.5ex}Educational Goals \\
\midrule
\end{tabularx}

\begin{tabularx}{\textwidth}{@{}YYYY@{}}
Identify the direct and indirect stakeholders. &
Identify the learning context. &
Identify the main responsible AI principles. &
Identify the educational goals. \\
\midrule
\end{tabularx}

\noindent
\begin{tabularx}{\textwidth}{@{}>{\columncolor{white}\centering\bfseries\arraybackslash}X@{}}
\rule{0pt}{2.5ex}Perceptions \\
\end{tabularx}

\noindent\rule{\textwidth}{0.4pt}

\begin{tabularx}{\textwidth}{@{}XXXX@{}}
Attributes & User Experience & Impact on Process / Outcome & Ethical \& Societal Implications \\
\bottomrule
\end{tabularx}
\end{table*}

\subsection{Prioritizing perceptions via evaluation rubric}
\label{sec:putting:rubric}

After mapping the learning environment, the workflow transitions to evaluating the relevance of stakeholder perceptions. In responsible AI, rubrics or matrices are commonly used to operationalize abstract framework elements~\cite{homewood2025third,
wang2025putting}. In \framework, this phase is supported by a comprehensive baseline rubric summarized in Table~\ref{tab:stakeholder_rubrics}. This rubric intersects nine distinct stakeholder groups with four core categories from our perception taxonomy: Attributes, User Experience, Impact on Processes/Outcomes, and Ethical and Societal Implications. 

Each intersection within the matrix is evaluated as having high, medium, or low relevance based on a synthesized review of standard responsible AI literature and operational field realities. Direct classroom participants, such as students and teachers, naturally score highest across these dimensions due to their proximity to the models, whereas policymakers score lower, with relevance concentrated primarily in ethical implications. By cross-referencing their initial situational map with this reference rubric, users can bypass exhaustive, generalized lists of AI concerns and instantly identify the cognitive or ethical friction points that require immediate attention.

\begin{table*}[h!]
\centering
\small 
\caption{Summary of stakeholder perceptions across four rubrics}
\label{tab:stakeholder_rubrics}
\begin{tabularx}{\textwidth}{>{\raggedright\arraybackslash}p{2.5cm} X X X X}
\toprule
\textbf{Stakeholder} & \textbf{Attributes} & \textbf{User Experience} & \textbf{Impact on Process / Outcome} & \textbf{Ethical and Societal Implications} \\
\midrule
Students & \cellcolor{highcolor}\textbf{High.} Accuracy and technical specifications ensure AI tools support learning effectively. & \cellcolor{highcolor}\textbf{High.} Ease of use and perceived usefulness are critical for student engagement. & \cellcolor{highcolor}\textbf{High.} Critical thinking and student motivation directly shape learning outcomes. & \cellcolor{highcolor}\textbf{High.} Privacy and academic integrity are vital due to students’ direct interaction with AI. \\
\midrule
Teachers & \cellcolor{highcolor}\textbf{High.} Accuracy and knowledge base are essential for reliable classroom support. & \cellcolor{highcolor}\textbf{High.} Ease of use enhances teachers’ ability to integrate AI into teaching. & \cellcolor{highcolor}\textbf{High.} Efficiency and critical thinking skills improve teaching and learning quality. & \cellcolor{highcolor}\textbf{High.} Fairness and job replacement concerns impact teachers’ adoption of AI tools. \\
\midrule
Parents & \cellcolor{mediumcolor}\textbf{Medium.} Parents value convenience but are less focused on technical details. & \cellcolor{highcolor}\textbf{High.} Perceived usefulness drives parents’ trust in AI tools for their children. & \cellcolor{mediumcolor}\textbf{Medium.} Parents care about student motivation but are less involved in daily outcomes. & \cellcolor{highcolor}\textbf{High.} Privacy and social justice are key to ensure equitable benefits for their children. \\
\midrule
School Administrators & \cellcolor{mediumcolor}\textbf{Medium.} Administrators prioritize functionality over specific technical attributes. & \cellcolor{mediumcolor}\textbf{Medium.} Perceived usefulness ensures AI aligns with school-wide goals. & \cellcolor{highcolor}\textbf{High.} Cost-effectiveness and student performance justify AI adoption at scale. & \cellcolor{highcolor}\textbf{High.} Social justice ensures equitable implementation across diverse student populations. \\
\midrule
System Owners & \cellcolor{mediumcolor}\textbf{Medium.} Owners ensure basic functionality but leave detailed specs to developers. & \cellcolor{mediumcolor}\textbf{Medium.} Perceived usefulness gauges overall adoption success in educational systems. & \cellcolor{highcolor}\textbf{High.} Efficiency and cost-effectiveness drive scalability of AI solutions. & \cellcolor{highcolor}\textbf{High.} Ethical concerns like privacy are prioritized to meet regulatory standards. \\
\midrule
Ed-tech Professionals & \cellcolor{highcolor}\textbf{High.} Knowledge base and accuracy are crucial for designing effective AI tools. & \cellcolor{mediumcolor}\textbf{Medium.} Ease of use supports deployment, but their focus is on technical integration. & \cellcolor{mediumcolor}\textbf{Medium.} Efficiency in implementation is relevant, though outcomes are less direct. & \cellcolor{mediumcolor}\textbf{Medium.} Privacy is noted, but their role emphasizes technical compliance over ethics. \\
\midrule
AI Developers & \cellcolor{highcolor}\textbf{High.} Accuracy and technical specs ensure AI tools meet educational needs. & \cellcolor{mediumcolor}\textbf{Medium.} Developers consider ease of use but prioritize functionality over experience. & \cellcolor{mediumcolor}\textbf{Medium.} Performance expectancy validates the tool’s impact on educational goals. & \cellcolor{mediumcolor}\textbf{Medium.} Developers address fairness but focus less on broader societal implications. \\
\midrule
Policy Makers & \cellcolor{lowcolor}\textbf{Low.} Policy makers are detached from specific technical attributes of AI tools. & \cellcolor{lowcolor}\textbf{Low.} They focus on policy alignment rather than daily user experience. & \cellcolor{mediumcolor}\textbf{Medium.} Performance expectancy shapes policies on AI’s role in education. & \cellcolor{highcolor}\textbf{High.} Privacy and fairness are critical to develop equitable AI regulations. \\
\bottomrule
\end{tabularx}
\end{table*}

\subsection{Application of the framework}
\label{sec:putting:application}

In the final step, the fully mapped and labeled matrix transforms into an active decision artifact that guides the responsible integration of LLM-based tools. To demonstrate this application stage, we examine two case studies.

\smallskip\textbf{Case 1: AI teaching assistant and tutoring pilot in elementary school.} This scenario describes an active pilot program involving an LLM-based teaching assistant and tutoring system in an elementary school setting managed by the New York City Department of Education. 

The implementation team initiated the \framework workflow by mapping the learning situation, summarized for a subset of stakeholders in Table~\ref{tab:case1_mapping}. The team identified K--5 students and teachers as direct stakeholders, parents as critical indirect guardians, and the context as primary school STEM tutoring, aiming to ease extreme teacher workloads. 

Next, the team applied the reference rubric to prioritize perceptions across these groups, flagging child-appropriate user experience constraints and the risk of automated overreliance as high-relevance vulnerabilities for students while marking data privacy and algorithmic transparency as high-priority ethical concerns for parents. This step surfaced a \emph{structural tension} that a top-down adoption process would likely have obscured: the institutional goal of reducing teacher workload pulls toward broad, unsupervised AI access, while the high-relevance overreliance risk for K--5 students pulls in the opposite direction. A procurement-driven process centered on the NYC DOE and the ed-tech vendor would have optimized for the former, with student and parental concerns surfacing only after deployment, if at all.  

In the final phase, this prioritized assessment directly guided institutional policy: the NYC DOE restricted the LLM to synchronous classroom hours under direct teacher supervision to mitigate student overreliance and enforced strict data-minimization mandates on the edtech vendor to address parental privacy concerns.

\begin{table*}[t!]
\centering
\small 
\caption{\framework mapping for Case 1: AI Teaching Assistant and Tutoring Pilot in Elementary School}
\label{tab:case1_mapping}

\begin{tabularx}{\textwidth}{@{}>{\columncolor{lightpurple}\bfseries}Y
                            >{\columncolor{lightgrey}\bfseries}Y
                            >{\columncolor{lightblue}\bfseries}Y
                            >{\columncolor{lightblue}\bfseries}Y@{}}
\toprule
Stakeholder & Context & R-AI Goals & Educational Goals \\
\midrule
\end{tabularx}

\begin{tabularx}{\textwidth}{@{}C{3.5cm}YYY@{}}
\textbf{Teachers} &
Daily classroom use of AI assistant in K-5 education &
Fairness, accountability, transparency; data privacy &
Reduce workload; improve learning outcomes \\
\midrule
\end{tabularx}

\begin{tabularx}{\textwidth}{@{}>{\columncolor{white}}Z
                            >{\columncolor{highcolor}}X
                            >{\columncolor{highcolor}}X
                            >{\columncolor{highcolor}}X
                            >{\columncolor{highcolor}}X@{}}
\rotatebox{90}{\textbf{Perceptions}} &
\textbf{Attributes.} Reliability and accuracy critical for trust. &
\textbf{User Experience.} Usability impacts integration into teaching. &
\textbf{Impact on Process / Outcome.} Affects instructional time and management. &
\textbf{Ethical and Societal Implications.} Concerns about fairness and data use. \\
\midrule
\end{tabularx}

\begin{tabularx}{\textwidth}{@{}C{3.5cm}YYY@{}}
\textbf{Students} &
Direct recipients of AI tutoring and support in math &
Privacy and autonomy; fairness in algorithmic support &
Personalized learning; improve comprehension and confidence \\
\midrule
\end{tabularx}

\begin{tabularx}{\textwidth}{@{}>{\columncolor{white}}Z
                            >{\columncolor{highcolor}}X
                            >{\columncolor{highcolor}}X
                            >{\columncolor{highcolor}}X
                            >{\columncolor{highcolor}}X@{}}
\rotatebox{90}{\textbf{Perceptions}} &
\textbf{Attributes.} Accuracy and adaptability matter. &
\textbf{User Experience.} Child-appropriate interface needed. &
\textbf{Impact on Process / Outcome.} Direct influence on motivation and outcomes. &
\textbf{Ethical and Societal Implications.} Privacy and academic integrity key. \\
\midrule
\end{tabularx}

\begin{tabularx}{\textwidth}{@{}C{3.5cm}YYY@{}}
\textbf{Parents} &
Indirect representatives of children’s interests in school decisions &
Privacy protection; transparency in algorithm use &
Ensure effective, equitable learning and protect student well-being \\
\midrule
\end{tabularx}

\begin{tabularx}{\textwidth}{@{}>{\columncolor{white}}Z
                            >{\columncolor{mediumcolor}}X
                            >{\columncolor{highcolor}}X
                            >{\columncolor{mediumcolor}}X
                            >{\columncolor{highcolor}}X@{}}
\rotatebox{90}{\textbf{Perceptions}} &
\textbf{Attributes.} Expect baseline quality. &
\textbf{User Experience.} Trust shaped by clarity of purpose. &
\textbf{Impact on Process / Outcome.} Interest in support for learning gaps. &
\textbf{Ethical and Societal Implications.} Strong concerns about data use and fairness. \\
\midrule
\end{tabularx}

\begin{tabularx}{\textwidth}{@{}C{3.5cm}YYY@{}}
\textbf{School Admins} &
Oversight of ed-tech implementation across K--5 classrooms &
Compliance, equity, transparency, and accountability &
Achieve system-wide improvements in efficiency and student performance \\
\midrule
\end{tabularx}

\begin{tabularx}{\textwidth}{@{}>{\columncolor{white}}Z
                            >{\columncolor{mediumcolor}}X
                            >{\columncolor{mediumcolor}}X
                            >{\columncolor{highcolor}}X
                            >{\columncolor{highcolor}}X@{}}
\rotatebox{90}{\textbf{Perceptions}} &
\textbf{Attributes.} Focus on scalable functionality. &
\textbf{User Experience.} Concerned with operational usability. &
\textbf{Impact on Process / Outcome.} Impacts teaching models, resource allocation. &
\textbf{Ethical and Societal Implications.} Responsible use across diverse populations. \\
\bottomrule
\end{tabularx}
\end{table*}

\smallskip\textbf{Case 2: AI curriculum support tool in higher education.} A similar trajectory emerges when applying the workflow to a higher education scenario involving an AI-driven curriculum support tool used by a social studies professor in Chicago. As shown in Table~\ref{tab:case2_mapping}, the instructor began by documenting the situational parameters and identifying adult college students and university faculty members as the primary stakeholders operating within a humanities setting to maximize instructional efficiency and generate differentiated course materials. 

In the prioritization stage, the instructor used the rubric to surface tensions: grading accuracy and perceived fairness of automated evaluations were students' primary concerns, while faculty prioritized technical reliability and academic integrity. This revealed \emph{a tension a conventional top-down process would have suppressed}: the capabilities that maximize instructional efficiency for faculty (\eg automated rubric generation, formative feedback) register as fairness and accuracy concerns for the students receiving evaluations. In a humanities context where evaluation is inherently interpretive, a faculty-led adoption decision without structured stakeholder mapping would likely have treated student concerns as an afterthought.

The final application step transformed this structured mapping into an explicit classroom syllabus: the AI tool was authorized to generate formative task outlines and peer-review scaffolding to maximize instructional efficiency; however, all summative evaluations remained manual, and students were fully transparent about how models were used to generate course prompts to preserve academic integrity.

\section{Exploring stakeholders sensemaking}
\label{sec:validation}

This section explores how two stakeholder groups, university faculty and K-12 parents, make sense of \framework in practice. Rather than providing empirical validation or representative claims about stakeholder perceptions, the goal is to illustrate how the framework surfaces questions, tensions, and possible extensions when applied under conditions of uncertainty and evolving norms around AI use in education.

\subsection{Methods}
\label{sec:validation:methods}

We secured IRB approval and conducted five in-person focus groups with university faculty ($N = 27$) across three large universities in North America and Europe, and three online focus groups with K-12 parents ($N = 24$) across the United States; all sessions were approved by the ethical review boards of our respective institutions. Both groups followed the same protocol: participants completed a brief background questionnaire on their familiarity with AI and responsible AI concepts, viewed a short video introducing \framework, and engaged in structured discussion about their experiences with AI in educational settings and their reflections on the framework's usefulness, limitations, and implications. Sessions lasted about one hour and were audio-recorded following informed consent; recordings were transcribed and analyzed using qualitative content analysis with an iterative codebook combining deductive codes informed by \framework dimensions and inductive codes that emerged from the data. See Appendix~\ref{app:focus_groups} for details.

\subsection{Findings}
\label{sec:validation:findings}

Faculty and parents raised remarkably similar concerns related to stakeholder boundaries, contextual complexity, and policy fragmentation, alongside group-specific insights that reflect each group's distinct relationship to AI adoption in schools. The themes below illustrate how applying \framework surfaces tensions, blind spots, and value conflicts that are difficult to anticipate from abstract principles alone, and are not intended to generalize across educational contexts.

\smallskip\textbf{Expanding the boundaries of relevant stakeholders.} Both faculty and parents reasoned about stakeholders whose interests are indirectly affected by classroom AI use but are rarely represented in formal decision-making. Faculty participants raised concerns about individuals whose personal information might be inadvertently shared with LLMs, as well as employers who shape curricular expectations: \emph{``I'm thinking about those kind of stakeholders, which aren't in the room, they're not even in the conversation, but they might be your roommates, because you've entered your address onto your CV, maybe accidentally [P4].''} Parents similarly emphasized overlooked groups, particularly children with developmental disabilities: \emph{``Children who have disabilities are kind of being left out [P22].''} Both groups also pointed to the broader community as a stakeholder whose interests extend beyond immediate educational outcomes. These reflections illustrate how applying \framework prompts consideration of indirect stakeholders whose perspectives may otherwise remain unexamined.

\smallskip\textbf{Context extends beyond the classroom.} Participants in both groups emphasized that the context of AI use cannot be reduced to discipline or educational level alone. Faculty pointed to institutional factors such as course size, whether a course is required or elective, and the broader messaging students receive about AI: \emph{``It's not just about what happens in the classroom. It's about
orientation, advising, and what students are being told this technology is for [P17].''} Parents highlighted teacher-related contextual factors, including teachers' readiness to adopt technology and their capacity to provide emotional and instructional support, as well as the differential needs of students with disabilities. These observations suggest that effective use of \framework requires attending to the full ecology of an educational setting, not only its formal instructional parameters.

\smallskip\textbf{Policy fragmentation and inconsistent guidance.} Faculty and parents alike described navigating inconsistent guidance on AI use. Faculty reported that approaches within the same program ranged from outright prohibition to full allowance: \emph{``We have some courses saying you use it because of the topic, and others saying absolutely do not use this, or it counts as plagiarism. No wonder students are confused when there's no unified policy [P12].''} Parents echoed this concern from outside the institution, with many reporting being informed of AI adoption decisions only after the fact, or not at all: \emph{``I think the decision has been made already before onboarding the parents [P9].''} These reflections reinforce the need for tools that support structured, context-aware reasoning rather than prescriptive rules.

\smallskip\textbf{Difficulty operationalizing responsible AI principles.} Faculty expressed particular uncertainty about how to operationalize responsible AI principles such as fairness, accountability, and privacy in everyday teaching practice. Challenges were most acute around grading and assessment: \emph{``Things that I find myself struggling with are fairness and accountability, especially around grading. Sometimes people that use the LLM will have a better outcome. How do I ensure that's fair, or distinguish between different kinds of AI use [P9]?''} These reflections suggest that the value of \framework lies not in prescribing solutions, but in supporting structured reasoning about trade-offs and constraints that practitioners already face.

\smallskip\textbf{Educational goals and employability pressure.} Faculty discussions surfaced a recurring tension between preparing students for the labor market and preserving the pedagogical integrity of higher education. Some viewed AI-related skills as essential, while others resisted framing education primarily around employability: \emph{``Software development is not going to look the same anymore. Employers will expect students to know how to work with these systems [P7]''} versus \emph{``I want to teach critical thinking skills. I don't want to teach how to use AI systems [P10].''} Participants also noted broader pressures---including accreditation requirements and efficiency-driven institutional priorities---that can crowd out deeper learning objectives. These tensions surfaced limitations in how the framework's goals dimension was initially interpreted, motivating clearer distinctions between institutional motivations and pedagogical objectives.

\smallskip\textbf{Parental inclusion and decision-making.} A distinctive theme from the parent focus groups was the wide variation in parental involvement during AI tool implementation in schools, ranging from complete exclusion to active partnership: \emph{``My consent was also asked. And I gave my full consent, and I was carried along through the whole process of implementing [P21].''} Many parents, however, reported learning about AI adoption incidentally or after decisions had been finalized. Several called for a structural shift in how frameworks like \framework are deployed: \emph{``I feel parents should actually be moved from being informed after AI decisions have been made to being active partners in shaping how AI is actually used in school [P10].''} These findings highlight parental engagement as a process requirement, not an afterthought.

\smallskip\textbf{Need for structured dialogue and parent training.} Parents expressed a clear desire for training on LLM-based tools and for structured opportunities to discuss them---both to become more informed and to influence policy. Participants also reported that the focus group discussions themselves shifted their views, suggesting that \framework can serve not only as a diagnostic instrument but also as a facilitation tool for structured stakeholder dialogue. A few participants offered direct feedback on the framework's design, noting that a categorized structure would be more usable than a flat list of considerations---feedback that aligns with the rubric-based approach developed in Section~\ref{sec:putting}. Additional quotes and observations from parent participants are provided in Appendix~\ref{app:focus_groups}.

\section{Conclusion, Limitations, and Future Work}
\label{sec:conclusion}

LLM-based systems are increasingly piloted in educational settings through partnerships between private companies and school districts, yet structured engagement with affected stakeholders remains limited. We introduced \framework, a stakeholder-first, context-sensitive framework for reasoning about perceptions
of LLM-based tools in education, grounded in a diagnostic analysis of prior work, illustrated through contextually distinct scenarios, and examined through focus groups with university faculty and K--12 parents. These contributions support more deliberate decision-making about whether, where, and how LLM-based systems should be adopted in alignment with responsible AI principles and educational goals.

\smallskip\textbf{Limitations.} First, stakeholder perception categories were derived from qualitative content analysis guided by researcher judgment and a targeted rather than systematic literature review; selection and coding decisions may shape the resulting taxonomy. This is consistent with exploratory synthesis of heterogeneous literatures, and \framework was designed to be iterative, allowing categories to evolve over time.  Second, the focus groups in Section~\ref{sec:validation} are not intended as empirical validation of \framework or to generalize across contexts or populations. Instead, they offer insight into how university faculty and K--12 parents reason about the framework under uncertainty, limited institutional guidance, and evolving norms around AI use in education. Tensions raised by both groups, and concerns around governance, accountability, and equity, point to directions for refinement.

\smallskip\textbf{Future work.} Applying \framework with additional stakeholder groups, including students, administrators, and policymakers, would surface perspectives currently underrepresented in the literature. A shared repository of cross-context findings could support institutional learning and inform policy. Longitudinal work could track how perceptions shift as LLM-based tools become more embedded in practice, and the open design of \framework invites continued refinement as educational technologies and governance structures evolve. Beyond the research agenda, we intend to bring \framework and its findings into the policy conversations unfolding today in higher education and K--12, where decisions about AI adoption are actively being made, and where structured stakeholder engagement is most urgently needed.

\newpage 
\section*{Ethical Considerations Statement}

This work examines stakeholder perceptions of LLM-based systems in educational settings and proposes a framework to support responsible reasoning about their adoption. Ethical considerations were addressed at multiple stages of the research. The diagnostic literature analysis relied exclusively on previously published studies and did not involve new data collection from human subjects. 

The faculty focus groups were conducted with informed consent, and participation was voluntary. To protect participants’ privacy and reduce the risk of identification, institutional affiliations and other potentially identifying details have been anonymized, and quotations are reported without attribution beyond participant identifiers. The focus groups were designed to elicit reflections on framework applicability rather than to evaluate individual practices or institutional policies.

The \framework framework is intended as a deliberative and supportive tool, not as a prescriptive or automated decision-making system. It does not produce recommendations, rankings, or judgments about stakeholders, institutions, or technologies. Instead, it is designed to surface perspectives, tensions, and considerations that may otherwise be overlooked. We acknowledge the risk that frameworks of this kind could be misused to legitimize predetermined decisions or to create the appearance of stakeholder engagement without meaningful participation. We therefore emphasize that \framework should be applied transparently, iteratively, and in conjunction with genuine stakeholder involvement and institutional accountability mechanisms.

\bibliography{main}

\newpage 
\begin{appendix}
\section*{Supplementary Materials}

\section{Diagnostic Literature Review}
\label{sec:survey} 

This literature review supports the diagnostic analysis in Section~\ref{sec:related:perceptions} and informs the taxonomy of perceptions used in the \framework framework. We reviewed scholarly studies of stakeholder perceptions of AI-based systems in educational settings, focusing on identifying prevalent perceptions across key stakeholder groups, such as students, teachers, parents, and administrators. We addressed the following question: \textbf{RQ:} Who are the key stakeholders and what are their perceptions as identified in the literature?

\subsection{Methods}
\label{sec:survey:methods}

\subsubsection{Selection procedure}
\label{sec:survey:methods:selection} The literature review is expected to provide a foundation for understanding existing perceptions and addressing gaps in the literature. Therefore, instead of a systematic review, we used Google Scholar as a data set, searching for scientific articles containing the search strings (``Large language models'' OR ``LLMs'' OR ``LLM-based tutor'' OR ``LLM-powered chatbot'' OR ``ChatGPT'') AND (``perceptions'' OR ``perspectives'' OR ``attitudes'') in their title, abstract, keywords, or introduction. For the selection procedure, we limited the number of results to the first 100 articles, using Google Scholar's relevance sort order.  

Articles were included if they met the following criteria: (1) discussed LLM-powered systems in education; (2) addressed perceptions of education stakeholders, such as students, teachers, parents, school staff, or education-related professionals; (3) were based on experimental, quasi-experimental, pre-experimental, or non-experimental research (excluding commentaries, secondary data analyses, and literature reviews); and (4) were written in English.  Finally, the study must involve LLM-powered chatbots. Studies that only mentioned ``AI tool'' without specifying whether they referred to large language model-based technology or rule-based systems were excluded.   This resulted in the selection of $N=38$ full-text articles for analysis.

\subsubsection{Content analysis}
\label{sec:survey:methods:analysis} After an initial reading of the articles, we identified recurring patterns or themes and developed a codebook to capture them. The codebook was developed combining codes taken from articles that employed existing theoretical frameworks (deductive) and codes that emerged from the articles without available codebooks or that did not develop one (inductive). In the latter cases, we annotated the results and discussion sections of the articles. As for the level of implication allowed, we included both explicit references to concepts and those implied by words or phrases. We allowed for flexibility by adding codes throughout the iterative readings.

Hence, we coded for the frequency, that is, we examined the presence of a selected code in the articles. We identified whether each code or concept reflected negative perceptions or concerns, positive perceptions or opportunities, or neutral perceptions. Additionally, we identified the specific stakeholder group(s) from which each code or concept originated. The validity of the coding process was ensured through multiple rounds of coding, allowing for refinement and consistency across the analysis. 

\subsection{Findings}
\label{sec:survey:findings}

We now summarize the basic break-down of the $N=38$ articles included in our analysis. As shown in Figure~\ref{fig:stats} (left), the majority of the included articles had samples from Asia ($N = 21$), followed by Europe ($N = 20$), North America ($N = 18$), Oceania ($N = 5$), Latin America ($N = 4$), and Africa ($N = 4$). (Note that several articles included samples from multiple continents.) When the location of the participants was not specified, the location of the authors' university was selected as a proxy.  Despite this geographic diversity, the literature largely converges on a narrow set of stakeholders and contexts, motivating the cross-context analysis in the body of the paper.

Regarding the contexts of application of the LLM-based chatbots, Figure \ref{fig:stats} (center) illustrates that the most frequent educational context of the studies was STEM ($N = 19$, 50\%), followed by English-as-foreign-language (EFL) ($N = 5$, 13\%) and Humanities ($N = 1$). A considerable number of studies did not specify the educational context ($N = 13$, 34\%).

Table \ref{tab:design} presents a break-down by type of study design. Most of the articles employed non-experimental study designs. 13 studies used interviews or surveys with exclusively open-ended questions, most of which were semi-structured. 12 studies used mixed methods, including surveys with both open- and closed-ended questions, as well as less common methodologies, such as having participants draw a visual representation of their perception of ChatGPT followed by a survey~\cite{ding2023students}. Three studies used quantitative questionnaires. Three articles used quasi-experimental research designs, two of which used pre- and post-test methodology, and one, mixed methods. One study was pre-experimental, using a one-shot case study design. Among the experimental designs, the most common was mixed methods, followed by randomized controlled trials (RCT) and multi-arm randomized trials. 

Consistent with the framing in the body of the paper, this analysis is diagnostic rather than evaluative and is not intended to constitute a systematic review.

\begin{table*}[t!]
    \centering
    \caption{Number of papers per type of study design}
    \begin{tabular}{llc}
        \toprule
         \multicolumn{2}{l}{\textbf{Study type}} & \textbf{\# studies}  \\
        \hline  
    \multirow{3}{*}{Non-experimental} & Interviews or open-ended questions surveys  & 13 \\ 
                                      & Mixed-methods         & 12            \\ 
                                      & Questionnaires          & 3            \\  \hline 
    \multirow{2}{*}{Quasi-experimental} & Pre-test/Post-test     & 2  \\
                                        &  Mixed-methods        & 1      \\   \hline 
    Pre-Experimental & One-shot case study     & 1    \\  \hline 
    \multirow{3}{*}{Experimental} & Mixed-methods             & 3 \\
            & Randomized controlled trial    & 2 \\
            & Multi-arm randomized trial      & 1\\
    \bottomrule
    \end{tabular}
    \label{tab:design}
\end{table*}

\subsubsection{\textbf{RQ:} Key stakeholders and their perceptions}
\label{sec:survey:findings:peceptions}

As shown in Figure~\ref{fig:stats} (right), students were the most studied stakeholder group, appearing in 31 of the 38 (82\%) studies, 27 of which involved University-level students and 4 involved K-12 students. Teacher perceptions were analyzed in 12 (32\%) studies.  Additionally, perceptions of industry professionals were considered in 1 study, albeit without explicitly specifying whether these were ed-tech professionals or learners.  No studies included any other stakeholders, such as parents or school administrators.

We identified different types of perceptions: positive (opportunities/benefits), negative (concerns), and neutral. We organized these perceptions into four categories: \textit{Attributes}, \textit{Impact on Processes or Outcomes}, \textit{User Experience}, and \textit{Ethical and Societal Implications}.  

\begin{figure*}[t!]
\begin{subfigure}[h]{0.48\linewidth}
    \includegraphics[width=\linewidth]{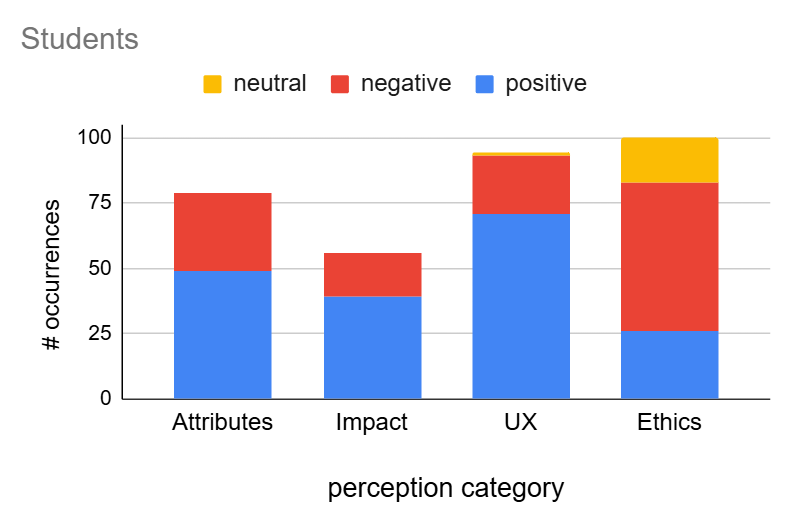}
    \caption{students}
    \label{fig:perceptions:students}
\end{subfigure}
\hfill 
\begin{subfigure}[h]{0.48\linewidth}
    \includegraphics[width=\linewidth]{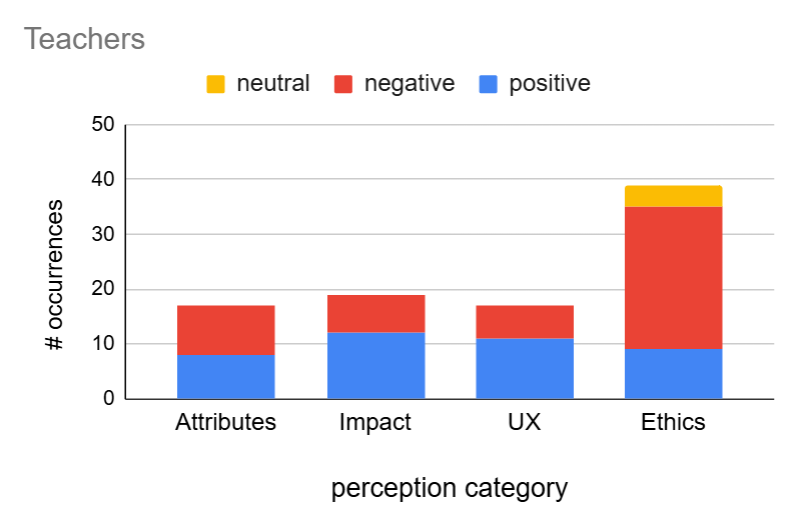}
    \caption{teachers}
    \label{fig:perceptions:teachers}
\end{subfigure}
\caption{[Summary of perceptions of two major stakeholders, students (\ref{fig:perceptions:students}) and teachers (\ref{fig:perceptions:teachers}), reported in the articles in our literature review.  Here, \emph{Impact} refers to ``Impacts on Processes and Outcomes,'' \emph{UX} refers to ``User Experience,'' and \emph{Ethics} refers to ``Ethical and Societal Implications.'' See Appendix~\ref{app:stats} for a detailed break-down. }
\label{fig:perceptions}
\end{figure*}

Figure~\ref{fig:perceptions} summarizes results of our analysis, showing the total number of mentions of perceptions in each category separately for two stakeholder groups, Students (Figure~\ref{fig:perceptions:students}) and Teachers (Figure~\ref{fig:perceptions:teachers}).   We present additional details about each perception category in the remainder of this section.   Note that, while we do not provide quantitative details about industry professionals as a stakeholder group, because they were only covered in one article, we do include them when discussing qualitative findings.

Overall, the perceptions of  students were more frequently positive than either negative or neutral, whereas teachers' were slightly more frequently negative: 56\% of all perceptions were positive and 38\% were negative for students vs. 43\% positive and 52\% negative for teachers.  Additionally, while both students' and teachers' perceptions focused heavily on the ethics of LLM-based systems, particularly in a negative or concerned light, students also expressed multiple views on UX, particularly emphasizing opportunities and positive aspects. In contrast, teachers expressed fewer opinions on UX as well as on other dimensions.

\paragraph{Attributes.} This category refers to perceptions of the characteristics of LLM-based chatbots. The papers we surveyed discussed the perceptions of students, teachers, and industry professionals regarding these attributes. The key attributes identified included: Convenience (also referred to as ``accessibility'', ``responsiveness'' or ``speed''), Accuracy (also referred to as ``confidence and trust in LLMs''), Knowledge base, Anthropomorphism (also referred to as ``human-like qualities'', ``perceived humanness'' and ``Natural Language Processing (NLP) capabilities''), and various other, less popular technical specifications, such as character limit and code editing. Perceptions of these attributes were mixed, with each being viewed both positively and negatively, except for Convenience, which was consistently seen as an advantage. Additionally, no neutral perceptions were found regarding these attributes. 

The most frequent positive perception among students and teachers is Convenience. For example, students in \citet{kazemitabaar2024codeaid} appreciated CodeAid's ``24/7 availability'', whereas teachers in \citet{mohamed2024exploring} highlighted the bot's ``real-time interaction and feedback''.

In contrast, the most recurring concern for students, teachers, and industry professionals is Accuracy. For example, students in \citet{weber2024measuring} were concerned with the accuracy of answers provided by LLMs, while teachers in \citet{ghimire2024generative} stated that generative AI can produce ``incorrect or fabricated results.'' However, an industry professional in \citet{chen2024learning} expressed that although she had ``low trust in ChatGPT'', she felt more confident after using it, as it helped her understand what she needed to figure out and made her in general faster. 

\paragraph{Impact on Processes and Outcomes.} This category refers to perceptions related to teaching and learning processes and their results. These perceptions were most commonly expressed by students and teachers, and include: Efficiency, Student motivation, Performance expectancy (a concept from UTAUT \cite{venkatesh2003user} and UTAUT2 \cite{acosta2024analysis}, 
referring to improved student learning and outcomes, Cost-effectiveness, as well as Critical thinking and problem solving skills. Among these, Performance expectancy, Cost-effectiveness, and Critical thinking and problem solving skills generated mixed perceptions. No neutral perceptions were found regarding these impacts.

Efficiency was the most commonly cited positive perception among students, teachers, and industry professionals. For example, students in \citet{kazemitabaar2024codeaid} stated that AI ``helps them work more efficiently,'' while teachers in \citet{jansson2024initial} reported that ``not having to calculate the problem themselves may free up time and make them more efficient as coaches, enabling them to help more students.'' Similarly, industry professionals in \citet{chen2024learning} noted that ``with LLMs, human time and energy could be liberated for more high-level tasks.''

The most common negative perception across students and teachers was the impact on Critical thinking and problem solving skills. For example, a student in \citet{woithe2023understanding} noted that while ChatGPT ``is a tool that's very useful when learning,'' they were ``still a bit scared [that] I'll become incompetent to think on my own.'' Teachers in \citet{ghimire2024generative} also expressed concern, stating that LLMs ``decrease critical thinking.'' 

\paragraph{User Experience (UX)} This category of perceptions relates to how individuals feel about interacting with LLM-powered systems. It includes concepts such as perceived Usefulness, Ease of use, Intention to use, Hedonic motivation, Well-being, Perceived importance, and Human interaction loss. Many of these—like  Usefulness, Ease of use, Intention to use, and Hedonic motivation—originated in the Technology Acceptance Model (TAM) \cite{davis1989perceived} and evolved in the Unified Theory of Acceptance and Use of Technology (UTAUT) \cite{venkatesh2003user} and UTAUT2 \cite{acosta2024analysis}, often with different names. 
For instance, perceived Usefulness in TAM is called Performance expectancy in UTAUT \cite{sair2018effect}, but we retained the term perceived Usefulness due to its use in studies beyond these frameworks. Similarly, TAM’s Ease of use evolved into UTAUT’s Effort expectancy, reflecting the role of technical skills in technology acceptance \cite{venkatesh2003user}. Except for Human interaction loss, which is perceived only negatively, and Hedonic motivation, viewed positively, negatively, and neutrally, all other concepts are perceived both positively and negatively.

The most commonly reported positive perception among students was perceived Usefulness. Findings in \citet{elkhodr2023ict} show ``a positive perception of ChatGPT as useful'', while \citet{lee2023learning} state that ``Learners’ perception of the AI Teaching Assistant (TA) is on par with that of human TAs in terms of [...] helpfulness.'' The most positive perception among teachers was Ease of use. For example, \citet{jansson2024initial} reports that ``The coaches expressed surprise about how friendly and supportive the AI bot was,'' while teachers in \citet{iqbal2022exploring} mentioned that ChatGPT is ``extremely user-friendly.'' Industry professionals reported positive perceptions of both Usefulness and Intention to use~\cite{chen2024learning}. 
 
The most common concern among both students and teachers is Human interaction loss. For example, a student in \citet{Ogbo-Gebhardt2024} disclosed that ``she felt a bit of anxiety considering how increased adoption of LLM tools could lead to increased social isolation,'' while \citet{chan2023students} reported that ``in academic institutions and education, some were concerned that the widespread use of AI might affect the student–teacher relationship, as students may be 'disappointed and lose respect for teachers.''' Teachers also expressed concerns about the ``lack of human connection'' \cite{mohamed2024exploring}. Interestingly, Ease of use was equally as concerning as Human interaction loss among teachers, with some mentioning that ``they found it difficult to use'' \cite{iqbal2022exploring} when discussing ChatGPT.

The only neutral UX perception in the literature is Hedonic motivation, a UTAUT2 concept referring to the fun or pleasure of using a technology. For instance, \citet{lieb2024student} observed that ``the neutral hedonic evaluation indicates that users had more neutral feelings about how interesting or exciting it was to use NewtBot.''

\paragraph{Ethical and Societal Implications.} This category covers the anticipated impact of introducing LLM-based tools in schools, including Social justice, Job replacement, Reliance, Ethical and privacy concerns, Academic integrity, Social influence (a construct from UTAUT that captures the impact of family and friends on technology acceptance), Workplace changes, and Learning changes. While all were viewed positively and negatively, the last four additionally had neutral perceptions.

The most frequently mentioned positive perception among students is Reliance on LLM-based bots. For instance, \citet{smith2024toward} noted that ``few students described behaviors that appeared to be intentional cheating (\ie 'generating complete answers'),'' demonstrating a healthy level of reliance on LLMs. Similarly, \citet{kazemitabaar2024codeaid} reported that ``although some students, like S14, felt that they 'over-relied on it too much rather than thinking,' many students displayed signs of self-regulation.'' This positive perception is also evident among industry professionals. A participant in \citet{chen2024learning} compared ChatGPT (GPT-4) and NetLogo, commenting that ChatGPT ``assumed what I wanted it to do, whereas this one makes you specify your assumptions.'' They preferred NetLogo Chat’s approach because ``it makes you think about the code more.'' For teachers, the most common positive perception was social justice, encompassing ``equity,'' ``accessibility,'' and ``cultural understanding.'' Teachers in \citet{mohamed2024exploring} noted that ChatGPT could become a ``tool for enhancing language exposure and cultural understanding […] and a tool for enhancing language accessibility and equity.'' Similarly, teachers in \citet{lau2023ban} stated that ``AI may improve equity and access.''

The most frequent concern among students was also related to Reliance, which, in this case, takes the shape of overreliance. This concern is reflected in statements such as, ``I feel that everyone might gradually lose their ability to think, as people won’t be willing to think anymore. They’ll just rely on machines to do the thinking for them'' \cite{zhang2024future}, or ``While students are specifically informed that AI can make mistakes and the exam keys are even provided, almost half of the students still agree with all the answers provided by ChatGPT, regardless of whether the answers are correct or incorrect'' \cite{ding2023students}. Similar concerns about Reliance were also shared by industry professionals. The most common concern among teachers is Academic integrity, often framed as ``plagiarism'' or ``AI-assisted cheating.'' For example, teachers in \cite{ghimire2024generative} expressed concern about ``focus on product over process,'' while \citeauthor{iqbal2022exploring} report that most teachers surveyed ``found ChatGPT would be used by students for cheating, and it will make them lazy.''

Finally, the most common neutral perceptions regarding the Ethical and Societal Implications of LLM-powered tools were related to Changes in the workplace for students and both Changes in the workplace and Changes in learning for teachers. Regarding the former, \citet{rogers2024attitudes} found that ``Almost all students agreed that ChatGPT will be integral to their careers,'' while \citeauthor{bernabei2023students} reported that ``students recognized the practical relevance of using such tools in a university context, as they believed it would better prepare them for the future working world, where AI is anticipated to play a prominent role.'' As for the latter, a teacher in \cite{mohamed2024exploring} stated, ``I see ChatGPT playing an increasingly central role in EFL education in the future.''

\subsection{Supplementary evidence for diagnostic gaps}
\label{survey:gaps}

This section provides additional evidence supporting the diagnostic gaps identified in Section~\ref{sec:related:perceptions}. After reviewing relevant academic work on stakeholder perceptions of AI-based systems in education, we identified the following gaps:

\paragraph{Stakeholder representation.} Our literature review highlighted a lack of representation of stakeholders beyond students and teachers, such as parents and broader school staff. Figure~\ref{fig:stats} shows that students appeared as the most recurring stakeholders, followed by teachers. This finding mirrors a conclusion from a literature review on robot tutors by~\citet{smakman2020robot}, who also noted the lack of parental representation. It further underscores the need identified by \citet{louie2022designing} to include ``key stakeholders (students, parents, and teachers) as essential to ensure a culturally responsive robot.'' Finally, it echoes the concerns raised by \citet{zeide2019robot} that ``school procurement and implementation of these systems are rarely part of public discussion.''

\paragraph{Context representation.} The second gap highlighted by our literature review was the inadequate representation of the educational context, particularly in terms of discipline and educational levels of the students' sample. Figure~\ref{fig:stats} illustrates the contexts of application of LLM-based systems, with the most frequent ones being STEM, followed by studies with unspecified contexts. Additionally, a lack of diversity in educational levels could be observed in Figure~\ref{fig:stats}, where 27 out of the 31 studies involving students focusing on university-level students and only four studies involving K-12 students. This is a critical issue because, while university students can independently choose whether to use AI technologies, K-12 students typically do not have a say in the implementation of these technology within their curricula \cite{Singer.2023, School}. This underscores the need for further research on perceptions across a wider range of educational contexts.

\paragraph{Homogeneity in assessment of perceptions.} Finally, the wide variety of theories, scales, and terms used across the literature to assess and name perceptions highlights the need for standardization in these aspects. For example, in \citet{Ogbo-Gebhardt2024}, the theme of trust in and accuracy of the responses generated by LLM-based bots is considered a Facilitating condition (UTAUT), whereas in multiple other papers \cite{kazemitabaar2024codeaid, ding2023students, yilmaz2023augmented} it is treated as a separate concept. This lack of unified terminology makes it difficult to develop a clear understanding of how stakeholders perceive LLM-based technology. 

Building on the works described above, we advocate for the inclusion and systematic collection of stakeholder perspectives to inform the design and deployment of LLM-powered systems in the classroom.

\section{Supplementary statistics}
\label{app:stats}

Complete break-down of perceptions, for all perception types in our taxonomy, for two main stakeholder groups, namely, students and teachers, are presented in Table~\ref{tab:perceptions_summary}.

\begin{table*}[t!]
\centering
\small
\caption{Frequency of positive, negative, and neutral perceptions of AI tutors by students and teachers}
\label{tab:perceptions_summary}
\begin{tabularx}{\textwidth}{@{}l *{6}{>{\centering\arraybackslash}X}@{}}
\toprule
\textbf{Perception} & \textbf{Students (Positive)} & \textbf{Teachers (Positive)} & \textbf{Students (Negative)} & \textbf{Teachers (Negative)} & \textbf{Students (Neutral)} & \textbf{Teachers (Neutral)} \\
\midrule
\multicolumn{7}{l}{\textit{Attributes}} \\
Convenience            & 16 & 3 & 0 & 0 & 0 & 0 \\
Accuracy               & 14 & 2 & 20 & 6 & 0 & 0 \\
Knowledge base         & 7  & 2 & 4  & 0 & 0 & 0 \\
Anthropomorphism       & 9  & 1 & 4  & 1 & 0 & 0 \\
Other technical specs  & 3  & 0 & 2  & 2 & 0 & 0 \\
\textbf{Total (Attributes)} & 49 & 8 & 30 & 9 & 0 & 0 \\
\midrule
\multicolumn{7}{l}{\textit{Impact on Processes / Outcomes}} \\
Efficiency                  & 17 & 6 & 0 & 0 & 0 & 0 \\
Performance expectancy     & 11 & 1 & 1 & 0 & 0 & 0 \\
Student motivation          & 5  & 2 & 0 & 0 & 0 & 0 \\
Critical thinking \& problem solving & 6 & 2 & 16 & 6 & 0 & 0 \\
Cost-effectiveness          & 0  & 1 & 0 & 1 & 0 & 0 \\
\textbf{Total (Impact)}     & 39 & 12 & 17 & 7 & 0 & 0 \\
\midrule
\multicolumn{7}{l}{\textit{User Experience}} \\
Ease of use             & 16 & 4 & 3 & 2 & 0 & 0 \\
Intention to use        & 15 & 2 & 3 & 1 & 0 & 0 \\
Usefulness              & 19 & 3 & 4 & 0 & 0 & 0 \\
Hedonic motivation      & 12 & 1 & 5 & 1 & 1 & 0 \\
Well-being              & 8  & 1 & 2 & 0 & 0 & 0 \\
Perceived importance    & 1  & 0 & 0 & 0 & 0 & 0 \\
Human interaction loss  & 0  & 0 & 5 & 2 & 0 & 0 \\
\textbf{Total (UX)}     & 71 & 11 & 22 & 6 & 1 & 0 \\
\midrule
\multicolumn{7}{l}{\textit{Ethical \& Societal Implications}} \\
Social justice            & 6  & 4 & 6  & 5  & 0 & 0 \\
Job replacement           & 3  & 2 & 8  & 3  & 0 & 0 \\
Reliance                  & 8  & 0 & 14 & 6  & 0 & 0 \\
Ethical/privacy concerns  & 2  & 1 & 12 & 4  & 0 & 0 \\
Academic integrity        & 3  & 1 & 12 & 8  & 3 & 0 \\
Social influence          & 4  & 1 & 5  & 0  & 1 & 0 \\
Changes in workplace      & 0  & 0 & 0  & 0  & 9 & 2 \\
Changes in learning       & 0  & 0 & 0  & 0  & 4 & 2 \\
\textbf{Total (Ethics)}   & 26 & 9 & 57 & 26 & 17 & 4 \\
\midrule
\textbf{Grand Total}      & \textbf{185} & \textbf{40} & \textbf{126} & \textbf{48} & \textbf{18} & \textbf{4} \\
\bottomrule
\end{tabularx}
\end{table*}

\begin{table*}[t!]
\centering
\small 
\caption{\framework mapping for Case 2: AI Curriculum Support Tool in High School}
\label{tab:case2_mapping}

\begin{tabularx}{\textwidth}{@{}>{\columncolor{lightpurple}\bfseries}C{3.5cm}
                            >{\columncolor{lightgrey}\bfseries}Y
                            >{\columncolor{lightblue}\bfseries}Y
                            >{\columncolor{lightblue}\bfseries}Y@{}}
\toprule
Stakeholder & Context & R-AI Goals & Educational Goals \\
\midrule
\end{tabularx}

\begin{tabularx}{\textwidth}{@{}C{3.5cm}YYY@{}}
\textbf{Teachers} &
Pilot implementation in college curriculum  &
Fairness in grading, transparency, data security &
Improve instructional efficiency; support differentiated learning \\
\midrule
\end{tabularx}
\begin{tabularx}{\textwidth}{@{}>{\columncolor{lightyellow}}Z
                            >{\columncolor{highcolor}}X
                            >{\columncolor{highcolor}}X
                            >{\columncolor{highcolor}}X
                            >{\columncolor{highcolor}}X@{}}
\rotatebox{90}{\textbf{Perceptions}} &
\textbf{Attributes.} Accuracy and reliability of grading are essential. &
\textbf{User Experience.} Ease of use impacts integration into curriculum. &
\textbf{Impact on Process / Outcome.} Time savings and teaching support noted. &
\textbf{Ethical and Societal Implications.} Fairness and data privacy are key concerns. \\
\midrule
\end{tabularx}

\begin{tabularx}{\textwidth}{@{}C{3.5cm}YYY@{}}
\textbf{Students} &
Direct users of the tool’s outputs (assignments, grades, feedback) &
Privacy, fairness, and autonomy in AI-supported evaluation &
Receive effective, personalized, and fair instruction \\
\midrule
\end{tabularx}
\begin{tabularx}{\textwidth}{@{}>{\columncolor{lightyellow}}Z
                            >{\columncolor{highcolor}}X
                            >{\columncolor{highcolor}}X
                            >{\columncolor{highcolor}}X
                            >{\columncolor{highcolor}}X@{}}
\rotatebox{90}{\textbf{Perceptions}} &
\textbf{Attributes.} Accuracy and fairness in grading matter deeply. &
\textbf{User Experience.} Clarity and usefulness of feedback are important. &
\textbf{Impact on Process / Outcome.} Influences motivation and engagement. &
\textbf{Ethical and Societal Implications.} Privacy and transparency are top concerns. \\
\midrule
\end{tabularx}

\begin{tabularx}{\textwidth}{@{}C{3.5cm}YYY@{}}
\textbf{Parents} &
Indirect stakeholders in higher education &
Transparency in data use; fairness in learning outcomes &
Ensure academic quality; protect student well-being \\
\midrule
\end{tabularx}
\begin{tabularx}{\textwidth}{@{}>{\columncolor{lightyellow}}Z
                            >{\columncolor{mediumcolor}}X
                            >{\columncolor{highcolor}}X
                            >{\columncolor{mediumcolor}}X
                            >{\columncolor{highcolor}}X@{}}
\rotatebox{90}{\textbf{Perceptions}} &
\textbf{Attributes.} Expect transparency in system function. &
\textbf{User Experience.} Usability contributes to trust. &
\textbf{Impact on Process / Outcome.} Concerned with grading fairness. &
\textbf{Ethical and Societal Implications.} Strong concern for data use and fairness. \\
\midrule
\end{tabularx}

\begin{tabularx}{\textwidth}{@{}C{3.5cm}YYY@{}}
\textbf{School Administrators} &
Oversight of AI integration into school-wide curriculum delivery &
Compliance, equity, transparency, and oversight responsibilities &
Support instructional innovation; manage resources effectively \\
\midrule
\end{tabularx}
\begin{tabularx}{\textwidth}{@{}>{\columncolor{lightyellow}}Z
                            >{\columncolor{mediumcolor}}X
                            >{\columncolor{mediumcolor}}X
                            >{\columncolor{highcolor}}X
                            >{\columncolor{highcolor}}X@{}}
\rotatebox{90}{\textbf{Perceptions}} &
\textbf{Attributes.} Focus on implementation and reliability. &
\textbf{User Experience.} Relevant to monitoring usage and uptake. &
\textbf{Impact on Process / Outcome.} Links to teacher performance and student outcomes. &
\textbf{Ethical and Societal Implications.} Oversight of fairness and equity in deployment. \\
\midrule
\end{tabularx}

\begin{tabularx}{\textwidth}{@{}C{3.5cm}YYY@{}}
\textbf{Ed-tech Professionals} &
Design and deployment of curriculum support systems &
Accuracy, fairness, and technical compliance &
Enable scalable solutions for personalized learning \\
\midrule
\end{tabularx}
\begin{tabularx}{\textwidth}{@{}>{\columncolor{lightyellow}}Z
                            >{\columncolor{highcolor}}X
                            >{\columncolor{mediumcolor}}X
                            >{\columncolor{mediumcolor}}X
                            >{\columncolor{mediumcolor}}X@{}}
\rotatebox{90}{\textbf{Perceptions}} &
\textbf{Attributes.} Accuracy and specification requirements are critical. &
\textbf{User Experience.} Interface usability informs deployment success. &
\textbf{Impact on Process / Outcome.} Efficiency influences implementation. &
\textbf{Ethical and Societal Implications.} Compliance prioritized over broader ethics. \\
\bottomrule
\end{tabularx}
\end{table*}

\section{Models of Perception}
\label{app:models}

Complete breakdown of models and theories conceptualizing attitudes, beliefs, and/or behaviors toward technology, as they appear in the literature, to investigate stakeholders' perceptions of LLM-based chatbots. The breakdown supports the thesis outlined in the third gap identified in the literature, namely, the lack of homogeneity in the assessment of perceptions and the need for standardization in how perceptions are conceptualized.

These models illustrate the heterogeneity in how perceptions are conceptualized in the literature, supporting the third diagnostic gap identified in Section~\ref{sec:related:perceptions}.

\paragraph{Technology Acceptance Model (TAM)} was first developed in 1989 by \citeauthor{davis1989perceived} and \citeauthor{davis1989user}) as a tool to predict the likelihood of adopting a new technology within a group or an organization. Originating in the psychological theory of reasoned action and theory of planned behavior (TPB) \cite{kowalska2023diffusion}, TAM suggests that an individual's beliefs, attitudes, and intentions can explain their adoption and use of technology. TAM evaluates the influence of four internal variables on actual technology use, namely, Perceived Ease of Use (PEU); Perceived Usefulness (PU); Attitude Toward Use (ATU), and Behavioral Intention to Use (BI) \cite{davis1989perceived, turner2010does, kowalska2023diffusion}.

\paragraph{Unified Theory of Acceptance and Use of Technology (UTAUT)} originates as an extention of TAM and was originally used to explain user behavior and intentions in information system contexts \cite{venkatesh2012consumer, kowalska2023diffusion, shakib2023lecturers}. The UTAUT model comprises several key constructs that affect users’ acceptance and use of technology. These are: Performance Expectancy, Effort Expectancy, Social Influence and Facilitating Conditions. According to \citeauthor{venkatesh2003user}, the Effort Expectancy factor  derives from the Ease of Use factor included in the Technology Acceptance Model (TAM). Additionally, the UTAUT model includes four other moderating variables, i.e., Gender, Age, Experience, and Voluntarism of Use, that may influence the four main constructs \cite{venkatesh2012consumer, shakib2023lecturers}.

\paragraph{Extended Unified Theory of Acceptance and Use of Technology (UTAUT2)} is an extension of the prevision one, as it incorporates additional constructs such as hedonic motivation, price value, habit, and usage intention, besides the original ones \cite{acosta2024analysis}. Additionally, in UTAUT2, voluntarism of use was dropped as moderator. The predictive ability of UTAUT2 theory is found to be much higher in comparison to UTAUT \cite{tamilmani2021extended}. 

\paragraph{ABC Model of Attitudes}~\cite{eagly1998attitude} conceptualizes attitudes as being composed of three elements: Affect, Behavior, and Cognition. Affect refers to the individual’s feelings or emotional response toward an object. Behavior reflects the individual’s intentions or actions toward an object or situation. Cognition refers to the beliefs or thoughts an individual holds about an object \cite{ajlouni2023students}. These three components collectively constitute an individual’s attitude toward an object, person, issue, or situation. 

\paragraph{Cultural-Historical Activity Theory (CHAT)} offers a perspective for examining the intricate and dynamic systems in which AI is integrated into education. When CHAT is applied to technology or AI in education, it doesn't only consider the student-technology relationship, but the whole activity system (subject, object, tools, rules, community, and division of labour) within the wider socio-cultural and historical context \cite{carbonel2024emerging}. The underlying concept is that when a new element, such as a new technology, is introduced into the activity system, it often leads to tensions and contradictions. The model can help examine the system and begin envisioning new approaches. 

\paragraph{Situated Expectancy-Value Theory (SEVT),} renamed from the original expectancy-value theory, highlights that students' expectancy-value beliefs—how well they believe they will perform on an upcoming task—are situation-specific, meaning they vary across contexts and are influenced by situational characteristics \cite{eccles2020expectancy}. SEVT is built on two main components, subjective task value and expectation of success. Subjective task value  consists of four parts that influence achievement-related choices and performance:
\begin{itemize}
\item Attainment value: how important the task is to the individual.
\item Intrinsic value: how interesting or enjoyable the task is.
\item Utility value: how useful the task is to the individual.
\item Relative cost: the perceived cost of engaging in the task.
\item Expectation of success refers to how likely the individual believes they are to succeed at the task \cite{eccles2020expectancy}.
\end{itemize}

\paragraph{Extended three-tier Technology Use Model (3-TUM)} is based on the original 3-TUM, which integrates multidisciplinary perspectives—including motivation, social cognitive theory (SCT), theory of planned behavior (TPB), and the technology acceptance model (TAM)—to investigate attitudes toward information technology \cite{liaw2007understanding}. Individual attitudes are divided into three tiers: the individual characteristics and system quality tier, the affective and cognitive tier, and the behavioral intention tier. The individual characteristics tier includes factors such as self-efficacy, hedonic motivation, and self-regulation, while the system quality tier encompasses environmental factors. The affective and cognitive tier includes perceived satisfaction, perceived usefulness, and performance expectancy \cite{cai2023factors, liaw2008investigating}. The constructs of performance expectancy and hedonic motivation are inspired by the Extended Unified Theory of Acceptance and Use of Technology (UTAUT2). The extended model adds an additional tier: learning effectiveness \cite{cai2023factors}.

\paragraph{Mitcham’s philosophical framework of technology}~\cite{mitcham1994thinking}  presents a comprehensive view of technology, emphasizing that technological knowledge and volition, rooted in human nature, lead to technological activities and the creation of concrete technological objects. The framework accommodates the full historical and conceptual range of technology—from ancient to modern and from simple to complex forms—without restriction to any specific type. Using this framework, the Mitcham Score was developed to classify students’ descriptions of technology through four key elements, namely Objects, Activities, Knowledge and Volition \cite{svenningsson2020mitcham}.

\paragraph{Specific Forms of Digital Competence Needed to use ChatGPT}~\cite{xiao2023exploratory} is a conceptual model based on \cite{instefjord2017educating}'s work on the integration of professional digital competence in initial teacher education programmes. The framework comprises three main components: Technological proficiency (e.g., Be aware of the features of ChatGPT, Understand how ChatGPT works), Pedagogical compatibility (e.g., Think about and plan ways to use ChatGPT to enhance or transform language teaching and learning tasks, Implement tasks that use ChatGPT) and Social awareness (e.g., Have a critical awareness of the drawbacks of ChatGPT and consider them when planning and implementing tasks, Inform learners of the risks, ethical issues, and drawbacks of ChatGPT).

\section{Perception Scales}
\label{app:scales}

Breakdown of scales assessing attitudes, perceptions, feelings and/or behaviors toward technology as they appear in the reviewed studies. Similar to the breakdown of models, the present lists 
supports the thesis outlined in the third gap identified in the literature: the lack of homogeneity in the assessment of perceptions and the need for standardization in how perceptions are evaluated.

In addition to theoretical frameworks, several studies have assessed perceptions of LLM-based tutors using existing scales, either fully or partially: 

\paragraph{Relevance of Science Education (ROSE)}\cite{schreiner2004sowing} is a research instrument created as part of an international comparative project meant to shed light on students’ experiences, interests, priorities, images and perceptions that are relevant to the learning of science and technology and their attitudes towards these subjects. The questionnaire contains 247 questions across 10 sections and assesses students' interest in, attitude towards, and experiences in science and technology, as well as their opinion about environmental challenges and career aspirations.

\paragraph{Short Version of the User Experience Questionnaire (UEQ-S)} aims at collecting end-users' feelings, impressions, and attitudes that arise when experiencing a product. It consists of 26 items grouped into 6 scales, namely Attractiveness, Perspicuity, Efficiency, Dependability, Stimulation and Novelty \cite{schrepp2017design}. Schrepp, Hinderks, and Thomaschewski later developed a Short Version of the UEQ, reducing it to 8 items grouped into two scales, which measure two meta-dimensions: pragmatic and hedonic quality, while still covering the spectrum of product qualities assessed by the original UEQ.

\paragraph{Cognitive Load Scale (CLS)} is a ten-item inventory, developed and validated by Leppink et al. \cite{leppink2013development}. The scale is based on the cognitive load theory \cite{van2005cognitive} which posits that instructions or instructional material can impose three types of cognitive load on the learner: intrinsic load (IL), extraneous load (EL), and germane load (GL). Intrinsic load refers to the inherent difficulty of the instructional content, which is influenced by the learners' prior knowledge. Extraneous load is the unnecessary cognitive effort caused by poorly designed instruction. Germane load, introduced later in the theory, refers to the mental effort consciously invested by learners in processing the intrinsic load \cite{hadie2016assessing}.

\paragraph{Epistemically-Related Emotion Scales (EES)}\cite{pekrun2017measuring} measure multiple emotions during knowledge-generating or epistemic activities. The instrument constists of 7 three-item scales, which assess the emotions of surprise, curiosity, enjoyment, confusion, anxiety, frustration, and boredom.

\paragraph{Feedback Perceptions Questionnaire (FPQ)}\cite{strijbos2010peer} measures feedback perceptions, in educational contexts, in terms of perceived fairness, usefulness, acceptance, willingness to improve, and affect.

\paragraph{Questionnaire assessing Chinese secondary school students’ intention to learn AI} \cite{chai2020extended} draws on the Theory of planned behavior \cite{ajzen1991theory} and the TAM literature~\cite{davis1989perceived} to identify factors that may affect secondary students’ intention to learn AI. The survey involves nine factors: students’ knowledge about AI, general AI anxiety, and subjective norms are presented as background factors that influence their attitudes toward the behavior (i.e., perceived usefulness, social good, and attitude toward use), perceived behavioral control (confidence and optimism), and their behavioral intention to learn AI.

\paragraph{Attitude Toward Cheating (ATC) scale}\cite{gardner1988scale} predicts cheating behavior. Fourteen of the 34 items explicitly condemned cheating, while the remaining 20 items expressed tolerance toward cheaters. Since cheating is generally regarded negatively, students may be inclined to hide their true feelings. To promote objectivity and honesty, the items were worded without referencing the readers; that is, no items included personal pronouns. This approach ensured that the readers were not put on the defensive regarding their own behavior; instead, their attitudes were inferred from their judgments of others' behavior.

\paragraph{Pupils’Attitudes Towards Technology Short Questionnaire (PATT)}\cite{de1988techniek} has been recently reconstructed and revalidated by \citeauthor{ardies2013reconstructing}. The revision led to the creation of the PATT Short Questionnaire (PATT-SQ), where the original 58 items were reduced to 24, divided into six subfactors. These six subfactors are Career Aspirations, Interest in Technology, Tediousness of Technology, Positive Perception of Effects of Technology, Perception of Difficulty, and Perception of Technology as a Subject for Boys or for Both Boys and Girls. The PATT-SQ
was examined and further developed by \citeauthor{svenningsson2018understanding}, leading to the creation of the PATT-SQ-SE.

\paragraph{General Attitudes towards AI Scale (GAAIS)} was developed based on the rationale that older Technology Acceptance Scales, such as the TAM \cite{davis1989perceived}, reflect users’ individual choices to use technology, while AI often involves decisions made by others. The scale comprises 32 items divided into two subscales: one measuring positive attitudes (Positive GAAIS) and the other measuring negative attitudes (Negative GAAIS). The Positive GAAIS consists of questions related to utility, desired use, and emotions (such as excitement), while the Negative GAAIS contains items related to unethical uses, errors, and negative emotions associated with using AI technology.

\paragraph{Attitudes Towards Using ChatGPT (ATUC)}\cite{ajlouni2023students} is based on the ABC Model of Attitudes to evaluate undergraduate students’ attitudes toward ChatGPT as a learning tool. The ATUC comprises 22 items (6–26) distributed across three subscales: a) the affects subscale (comprising seven items, 6–12), b) the cognitive subscale 
(comprising eight items, 13–20), and c) the behavioral subscale (comprising seven items, 21–27).
\textbf{ Student Attitudes Toward Artificial Intelligence (SATAI) scale} The SATAI scale was developed by \citeauthor{suh2022development} and consists of 26 items that encompass three components—cognitive, affective, and behavioral factors—representing K-12 students' attitudes toward AI.

\paragraph{Computer Technology Use Scale (CTUS)}\cite{conrad2008relationships} includes five factors: computer efficacy, technology-related anxiety, complexity, positive attitudes, and negative attitudes. These factors are organized into three domains: computer efficacy (perceived capacity), attitudes toward technology, and technology-related anxiety.

\paragraph{Technology Acceptance Model Edited to Assess ChatGPT Adoption (TAME-ChatGPT)} survey was created based on the TAM framework~\cite{sallam2023assessing}. It comprised 13 items for participants who heard of ChatGPT but did not use it and 23 items for participants who used ChatGPT.

\section{Focus Group Details}
\label{app:focus_groups}

This section supplements the material in Section~\ref{sec:validation}.

\subsection{Protocol}

We conducted five in-person focus groups with university faculty ($N = 27$)
across three large universities in North America and Europe, and three online
focus groups with K-12 parents ($N = 24$) across the United States. All
sessions were approved by the ethical review boards of our respective
institutions prior to data collection.

Both groups followed the same structured protocol. At the start of each
session, participants completed a brief background questionnaire capturing
self-reported familiarity with AI and responsible AI concepts. They were then
shown a short video introducing \framework and its components. The subsequent
discussion centered on participants' experiences with AI in educational
settings and their reflections on the framework's usefulness, limitations, and
implications. Sessions lasted approximately one hour and were audio-recorded
following informed consent. The protocol allowed facilitators to ask follow-up
questions for clarification while keeping discussions participant-driven; no
consensus or ranking was required.

Recordings were transcribed and analyzed using qualitative content analysis.
We developed a codebook through an iterative process combining deductive codes
informed by \framework dimensions with inductive codes that emerged from the
data. Coding focused on recurring themes, points of uncertainty, and perceived
limitations or extensions of the framework. All coded segments were reviewed
by the full research team to ensure consistency and interpretive alignment.

\subsection{Participant characteristics}

Table~\ref{tab:faculty_participants} and Table~\ref{tab:parent_participants}
summarize the characteristics of faculty and parent participants, respectively.

\begin{table}[h]
\centering
\small
\caption{Faculty focus group participants ($N = 27$).}
\label{tab:faculty_participants}
\begin{tabular}{ll}
\toprule
\textbf{Characteristic} & \textbf{Value} \\
\midrule
Institution type      & Research universities \\
Geographic region     & North America, Europe \\
\midrule
\multicolumn{2}{l}{\textit{Disciplinary background}} \\
\quad Social Sciences & $N = 12$ \\
\quad STEM            & $N = 9$ \\
\quad Humanities      & $N = 6$ \\
\midrule
\multicolumn{2}{l}{\textit{Teaching role}} \\
\quad Professor          & $N = 26$ \\
\quad Assistant Professor & $N = 1$ \\
\midrule
AI use in teaching (1–5, low–high)         & Mean $= 3.00$ \\
Responsible AI familiarity (1–5, low–high) & Mean $= 3.70$ \\
\bottomrule
\end{tabular}
\end{table}

\begin{table}[h]
\centering
\small
\caption{Parent focus group participants ($N = 24$).}
\label{tab:parent_participants}
\begin{tabular}{ll}
\toprule
\textbf{Characteristic} & \textbf{Value} \\
\midrule
\multicolumn{2}{l}{\textit{Parent education level}} \\
\quad AA / AAS            & $N = 4$ \\
\quad BA / BS             & $N = 11$ \\
\quad MA / MS / MBA       & $N = 7$ \\
\quad PhD / EdD           & $N = 1$ \\
\quad Unknown             & $N = 1$ \\
\midrule
\multicolumn{2}{l}{\textit{Children's grade level (may overlap)}} \\
\quad Elementary school   & $N = 10$ \\
\quad Middle school       & $N = 13$ \\
\quad High school         & $N = 7$ \\
\midrule
AI use frequency (1–5, low–high)        & Mean $= 4.12$ \\
Self-reported AI understanding (1–5, low–high) & Mean $= 4.20$ \\
\bottomrule
\end{tabular}
\end{table}

\subsection{Additional quotes and observations}
\label{app:focus_groups:quotes}

The quotes below supplement those presented in Section~\ref{sec:validation},
organized by the themes reported in Section~\ref{sec:validation:findings}.

\subsubsection{Expanding the boundaries of relevant stakeholders.}

Faculty noted that employer expectations function as an indirect but powerful
force shaping student behavior and curricular priorities:
\textit{``Employers are already signaling what kinds of skills they expect,
and that shapes what students think matters, whether we like it or not
[P7].''}

Parents emphasized that decisions about AI in schools should foreground
family interests over those of commercial actors:
\textit{``Parents should always be the primary people that should have most
of the say. I don't think that companies should be involved. Reason being:
every company wants to make profits [P11].''}

\subsubsection{Context extends beyond the classroom.}

Faculty highlighted that course characteristics---particularly scale and
whether enrollment is required or elective---shape the stakes of AI adoption:
\textit{``I think the other component that I also use is the size. If it's
one of the required courses like my data ethics class, it's 81 students in
there, versus the smaller classes that are electives [P11].''}

Parents stressed teachers' readiness to adopt AI and the importance of human
oversight:
\textit{``I just want to have that strong belief that they [teachers] are not
totally depending on it, but are just using it as a tool.''}

\subsubsection{Policy fragmentation and inconsistent guidance.}

Parents described confusion arising from inconsistent school-level
communication:
\textit{``Schools should definitely define how AI be used, so students don't
get to misuse it. And now you know that it's a problem if students overuse it
or over-rely on it [P11].''}

Parents also raised cost and sustainability as practical concerns often absent
from policy discussions:
\textit{``I feel my major concern is about the cost and sustainability. Some
AI tools are really, really expensive, and schools may depend on tools they
can't maintain for a long period of time. The risk is just about inconsistency
in the learning support.''}

\subsubsection{Difficulty operationalizing responsible AI principles.}

Faculty described how accountability is often individualized and placed on
students in the absence of clear institutional guidance:
\textit{``Ultimately what you submit is your work. You take responsibility,
even if the AI helped you. That's how I try to make sense of accountability
[P7].''}

Parents raised data privacy as a primary concern, framing it in terms of
access and purpose limitation:
\textit{``Knowing who accesses my child's data is key, and the kind of
information they access should be strictly for their improvement.''}

\subsubsection{Educational goals and employability pressure.}

Faculty noted that student labor market anxiety translates into pressure on
instructors, even when employer expectations remain poorly defined:
\textit{``Students keep asking which skills will still matter and which
projects will be more lucrative. Companies don't clearly communicate what
those skills are [P9].''}

Participants also described how accreditation requirements and
efficiency-driven institutional priorities can crowd out deeper learning:
\textit{``We are dragged by the madness of accreditations and results. This
efficiency paradigm makes it hard to focus on deep learning [P21].''}

\subsubsection{Parental inclusion and decision-making.}

Parents who had experienced exclusion from the adoption process described
learning about AI use in their children's schools only incidentally:
\textit{``I kind of found out when my kid in high school wanted me to check
how well she did. That's when I found out, oh, it's allowed in your school
[P12].''}

Participants proposed concrete mechanisms for improving involvement, including
online information sessions and structured surveys:
\textit{``Probably short surveys where parents could attend meetings online
and probably give their own suggestions. Those would actually really make
sense.''}

\subsubsection{Need for structured dialogue and parent training.}

Several parents reported a shift in their views over the course of the focus
group itself, suggesting that Co-PALE can serve as a facilitation tool for
structured stakeholder dialogue:
\textit{``For instance, for me right now, I see [AI] in both positive and
negative light, because we can't ignore the bad just because there's some
good in it\ldots{} the positive is a little bit overriding the negative for
me, which at the beginning of the discussion, it was not [P3].''}

One participant offered direct usability feedback, noting that a categorized
structure would be preferable to a flat list of considerations---feedback
consistent with the rubric-based design in Section~\ref{sec:putting}:
\textit{``I think for me, this long list of questions is not very helpful.
What would be actually helpful is to categorize it. So there are different
categories here, like, in terms of helpfulness and usefulness, and the other
one is in terms of the educational value, the other one is about equity
[P2].''}

\end{appendix}

\end{document}